\documentclass{svjour3}                     
\smartqed  
\usepackage{graphicx}
\usepackage[authoryear]{natbib}
\usepackage{comment}
\usepackage{epstopdf}
\usepackage{subcaption}
\usepackage{multirow}
\usepackage{pifont}
\usepackage{amsmath}   
\begin{document}

\title{The Effects of Information Overload on Online Conversation Dynamics\thanks{This work was supported by DARPA program HR001117S0018  (FA8650-18-C-7823).}
}

\titlerunning{The Effects of Information Overload on Online Conversation Dynamics}        
\author{Chathika Gunaratne\textsuperscript{1}\and
Nisha Baral\textsuperscript{1}\and
William Rand\textsuperscript{2}\and 
Ivan Garibay\textsuperscript{1\dag}\and 
Chathura Jayalath\textsuperscript{1}\and 
Chathurani Senevirathna\textsuperscript{1}
}
\authorrunning{Gunaratne C. et al.}
%
\institute{\textsuperscript{1}University of Central Florida, Orlando, FL, 32816, USA \and
\textsuperscript{2}North Carolina State University, Raleigh, NC, 27695
\email{\dag igaribay@ucf.edu} 
}

\date{Received: date / Accepted: date}

\maketitle

\begin{abstract}
The inhibiting effects of information overload on the behavior of online social media users, can affect the population-level characteristics of information dissemination through online conversations. 
We introduce a mechanistic, agent-based model of information overload and investigate the effects of information overload threshold and rate of information loss on observed online phenomena.
We find that conversation volume and participation are lowest under high information overload thresholds and mid-range rates of information loss.
Calibrating the model to user responsiveness data on Twitter, we replicate and explain several observed phenomena: 
1) Responsiveness is sensitive to information overload threshold at high rates of information loss;
2) Information overload threshold and rate of information loss are Pareto-optimal and users may experience overload at inflows exceeding 30 notifications per hour; 
3) Local abundance of small cascades of modest global popularity and local scarcity of larger cascades of high global popularity explains why overloaded users receive, but do not respond to large, highly popular cascades; 
4) Users typically work with 7 notifications per hour;
5) Over-exposure to information can suppress the likelihood of response by overloading users, contrary to analogies to biologically-inspired viral spread.
Reconceptualizing information spread with the mechanisms of information overload creates a richer representation of online conversation dynamics, enabling a deeper understanding of how (dis)information is transmitted over social media.

\end{abstract}

\section{Introduction}
\label{intro}
Instant access to vast quantities of information has resulted in online social media users experiencing the effects of information overload \citep{gomez2014quantifying,hodas2013friendship,koroleva2010stop,li2014modeling,feng2015competing}. Information overload has been defined as the adverse state in which a decision maker's usual cognitive abilities are hindered by an excess of, or increased complexity in, received information (See \citep{roetzel2019information} for a comprehensive survey on information overload). Biological limits to cognition, in particular, memory and attention span, play an important role in information overload. In the case of asynchronous communication, besides the recollection of ideas and experiences required to maintain conversation, memory is also required to maintain a store of the conversations and individuals to whom responses are due. With the increasing dominance of online social media, asynchronous communication is quickly becoming the norm. Despite the typical preference of immediate response, studies have shown that most individuals are satisfied with accepting responses that are delayed up to a day \citep{morris2010people}. The downside of asynchronous communication is that it can quickly lead to a backlog of conversations demanding responses, which may exceed the individual's ability to respond, overwhelming them, and potentially driving them into a state of information overload. Although memory capacity, information overload, and the physiological stresses that result have been studied extensively in lab experiments \citep{cowan2001magical}, the effects of information overload on online conversation dynamics and the potential stresses it places on collective social behavior remain largely unstudied. 


Social media platforms employ automated filtering and recommendation algorithms in order to ease the flow of information to users and improve user engagement. Information is typically delivered through notification feeds such as Twitter's timeline or Reddit's subreddits. These platforms rapidly convey vast amounts of information consecutively, from multiple senders, to users, who would have otherwise taken considerably more time to perceive, process, and respond to through synchronous, direct forms of communication. As a result, users may be subjected to a constant state of information overload \citep{koroleva2010stop}. Gomez et al. discover that the responsiveness of Twitter users stays constant while they receive less than 30 incoming tweets an hour, yet beyond this limit, increasingly experience the effects of information overload \citep{gomez2014quantifying}. They discover that the reduction in responsiveness of an overloaded individual follows a power-law to the excess inflow of tweets received per hour. 
The interplay between limited attention span and the gradual reduction in visibility of incoming information over time has been shown to have significant effect on the selection of information to respond to, and in turn, the distance over which information is shared \citep{hodas2012visibility}. 


Information overload can have important implications on the spread of information online. \citep{feng2015competing} demonstrate that due to finite limits of human cognition, information diffusion may not necessarily follow the dynamics of SIR models of biological contagion as previously believed. Rather, when users are constantly bombarded with information, they are less likely to respond as their attention must be increasingly divided between multiple sources, in contrast to SIR models of disease, where a higher number of exposures leads to a higher chance of infection. Gon\c claves et al. demonstrate the importance of considering cognitive limits when modeling information diffusion, by showing how a simple, finite-sized memory queue in a probabilistic model of information diffusion was able to support Dunbar's thesis of biological limits to the number of social connections that may simultaneously be maintained by a single individual \citep{gonccalves2011modeling}. 

In this paper, we investigate the effects that information overload has on the properties of online conversations, such as information cascade volume, virality, and user participation.
We revisit several findings that are reported in the literature.
There is evidence that overloaded users typically receive larger, more popular cascades of information, but cannot be used as signalers for interesting information as they do not receive small developing cascades \citep{hodas2013friendship}. 
Li et al. provide a mathematical proof that, despite the intuition that users with larger \textit{view scope} (activity information stored in technological scaffolding) retain information longer and therefore have higher likelihood of response, the likelihood of response is unrelated to the size of one's view scope, because the larger the view scope of a user is the more messages get stored for processing and the lesser the likelihood is that one particular message gets responded to \citep{li2014modeling}. 
However, this proof assumes a fixed view scope size over time, regardless of the information overload experienced. 
Considering the findings of \citep{gomez2014quantifying}, we hypothesize that the inclusion of view scope suppression under information overload would cause the likelihood of response to be related to view scope size and match actual responsiveness data more closely.

We introduce a mechanistic model of information exchange over an asynchronous communication medium, both informed by results from analytical studies on online social media activity \citep{gomez2014quantifying, roetzel2019information} and motivated by concepts from extended-self theory \citep{belk2014digital} and working memory \citep{cowan2001magical}. With this model we analyze the effects of information overload on population-scale conversation properties and information dynamics.
While information received through direct, synchronous communication is stored in working memory \citep{baddeley2012working}, information received through asynchronous communication media, such as online social networking platforms, is first stored in a technological scaffolding, such as a notification feed, and selectively considered for action.
Accordingly, in the described model, \textit{messages} of activity of one's influencing neighbors that are available for processing are stored in an \textit{actionable information queue}, while the individual considers responding to this information. 
This queue is distinct from the complete stream of received information, which is archived on the technological scaffold, as the actionable information queue only represents the collection of information that an individual would instinctively respond to without an active search through the technological scaffolding of the communication medium. This information structure has also been referred to \textit{view scope} in previous mathematical models of information overload on social media \citep{li2014modeling}. 

\begin{figure*}[h!]
    \centering
    \includegraphics[width=\textwidth]{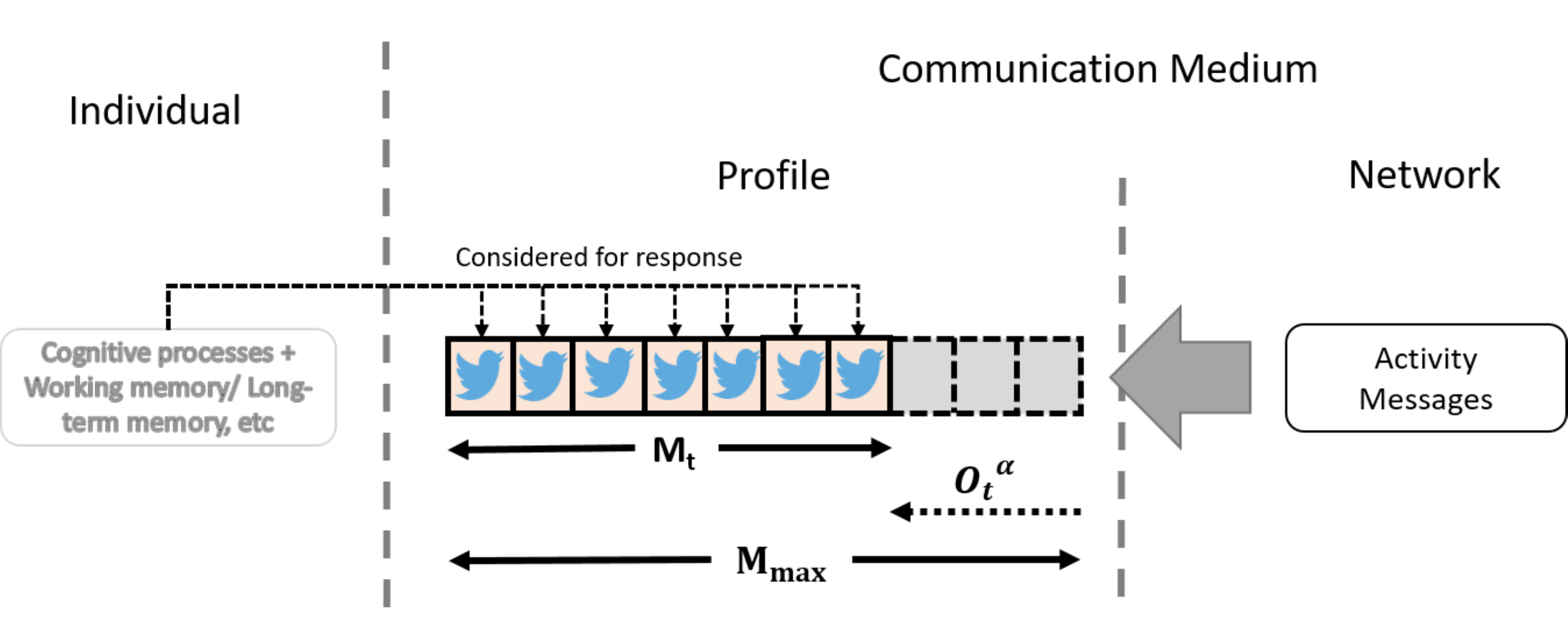}
    \caption{Conceptual depiction of information exchange via an asynchronous communication medium. Information is received into a temporary storage, which we term the \textit{actionable information queue} and considered for response by the user. When the amount of information received exceeds the \textit{information overload threshold} $M_{max}$, the \textit{current actionable information queue length}, $M_t$, is updated by subtracting the amount of excess information, $O_t$, raised to the power of the \textit{rate of information loss}, $\alpha$. }
    \label{fig:concept}
\end{figure*}

Fig. \ref{fig:concept} describes how this actionable information queue acts as a buffer for the incoming messages from the social network. We hypothesize that the maximum capacity of the actionable information queue is restricted to an \textit{information overload threshold}, $M_{max}$, by the biological and social limits of cognition; i.e. $M_{max}$ is the finite limit to the amount of information received from the communication medium that may be processed by an individual during a given time interval, without an active search of the archive of messages. Often, this rate of response is surpassed by the rate of incoming messages from the network, leading to an overflow of the actionable information queue. Under such information overload, individuals are subject to sub-optimal decision-making \citep{roetzel2019information}. Considering evidence from analytical studies \citep{gomez2014quantifying}, we calculate a new, sub-optimal actionable information queue size, or current \textit{actionable information capacity}, $M_t$. $M_t$ is calculated as the previous information capacity, $M_{t-1}$, minus, the amount of excessive information past $M_{max}$, $O_t$, raised to the power of a \textit{rate of information loss under overload} parameter, $\alpha$.

\section{Model Description}
In this section, we describe the model ensemble used to simulate information overload on asynchronous communication media. This ensemble comprises of the Multi-Action Cascade Model (MACM) of conversation \citep{gunaratne2019a}, upon which agents embodying the actionable information queue and information overload mechanism introduced above were modeled. Throughout the rest of the paper, the MACM and the information overload mechanisms were treated as a single ensemble model for sensitivity analysis, calibration, and explanatory simulations.

\label{sec:1}
\subsection{The Multi-Action Cascade Model}
\label{sec:1.1}
The MACM \citep{gunaratne2019a} derives from traditional diffusion of information models such as the independent cascade model \citep{granovetter1978threshold,bass1969new,Goldenberg2001,rand2015agent}, but it is unique, as it is the first of its kind to simulate diffusion of information in the form of conversations. The MACM follows the principles of conversation theory \citep{pask1976conversation} instead of merely simulating the binary adoption of a topic or opinion. The MACM is based on three premises: 

\begin{itemize}
    \item \textbf{Premise 1}: Diffusion of information over online social media occurs through conversations. Individuals participate in conversations due to the following factors: 1) influence of other participants, $Q$, 2) influence from information sources exogenous to the conversation, $P$, or 3) the internal need to participate in conversation, $I$.
    \item \textbf{Premise 2}: Conversation participants can perform three types of actions: 1) Initiation of a new conversation, 2) contribution to an existing conversation, 3) sharing existing information from a conversation.
    \item \textbf{Premise 3}: Given a particular topic of interest, the influences $q \in Q$, $p \in P$, and $i\in I$ can be determined from event timeseries data, by measuring the ratio of information flow, from the influencing timeseries to the influenced timeseries, to the total information flow of the influenced timeseries.  \citep{schreiber2000measuring, gunaratne2019a}.
\end{itemize}

MACM agents exist on a network of endogenous influence probabilities that govern the conditional probabilities that, given an agent's neighbor takes one of the three particular actions above, the agent is influenced to perform a particular action itself. When a MACM agent performs a particular action, they produce a message, representative of a social media notification, that indicates which user performed the action, the action type, and the conversation the action is being performed on. These messages are propagated to neighboring agents to whom the acting agent has a total influence probability greater than 0 over the receiving agent. The receiving agents then act on the incoming messages with a probability indicated by the influence probability the sender agent has over it, $q$. 

Once an MACM agent receives a message, it can then decide to perform one out of three actions: 1) \textit{initiate} a new conversation on the topic, 2) \textit{contribute} to the sender's conversation, or 3) \textit{share} the sender's conversation with other agents. For the experiments in this paper, values for $q$ between the agents are derived by considering all potential, directed, dyadic relationships between all users in the data provided. For each potential directed edge, the ratios of information flow over time from one user, as the influencer, to another, as the influenced, to the total information produced by the influenced agents is calculated. These conditional probabilities are calculated from user activity data as the ratio of the marginal transfer entropy, from the timeseries of the particular action of the influencer to the timeseries of action of the influenced, to the entropy of the particular action of the influenced. The advantage of this approach is that it does not assume that a retweet/reply structure must exist in order for influence to propagate, instead, every potential user-action to user-action timeseries is considered. As a result, influence that spans across retweet/reply/post structures and even across platforms can be estimated.
$p$ and $i$ may be calculated similarly, but for the purposes of the experiments in this paper, considering the endogenous influence probabilities $q$ is sufficient. The GPU-based implementation of the MACM in python can be found in \citep{gunaratne2019b}.

\subsection{Modeling Information Overload}
We model the mechanisms of information overload by integrating the \textit{actionable information queue} into agents modeled in the MACM. A further three premises are considered here:

\begin{itemize}
    \item \textbf{Premise 4}: Information exists in the form of countable units, as seen notification feeds of online social media platform \citep{gomez2014quantifying,li2014modeling,koroleva2010stop}.
    \item \textbf{Premise 5}: The total number of received messages that may be considered for response without experiencing overload in a given span of time is of a finite \textit{information overload threshold}, $M_{max}$ \citep{cowan2001magical,gomez2014quantifying,li2014modeling}.
    \item \textbf{Premise 6}: The effects of information overload are experienced once the number of incoming messages has exceeded the information overload threshold, $M_{max}$, after which a sub-optimal number of messages are considered for response at a given span of time, $M_{max} - M_t$. $M_t$, or the  current \textit{actionable information capacity}, follows a power-law relationship of exponent, $\alpha$, or  \textit{rate of information loss under overload}, with the amount of incoming information overload experienced \citep{gomez2014quantifying,koroleva2010stop,li2014modeling}.
\end{itemize}

\begin{figure*}[h!]
    \centering
    \includegraphics[width=\textwidth]{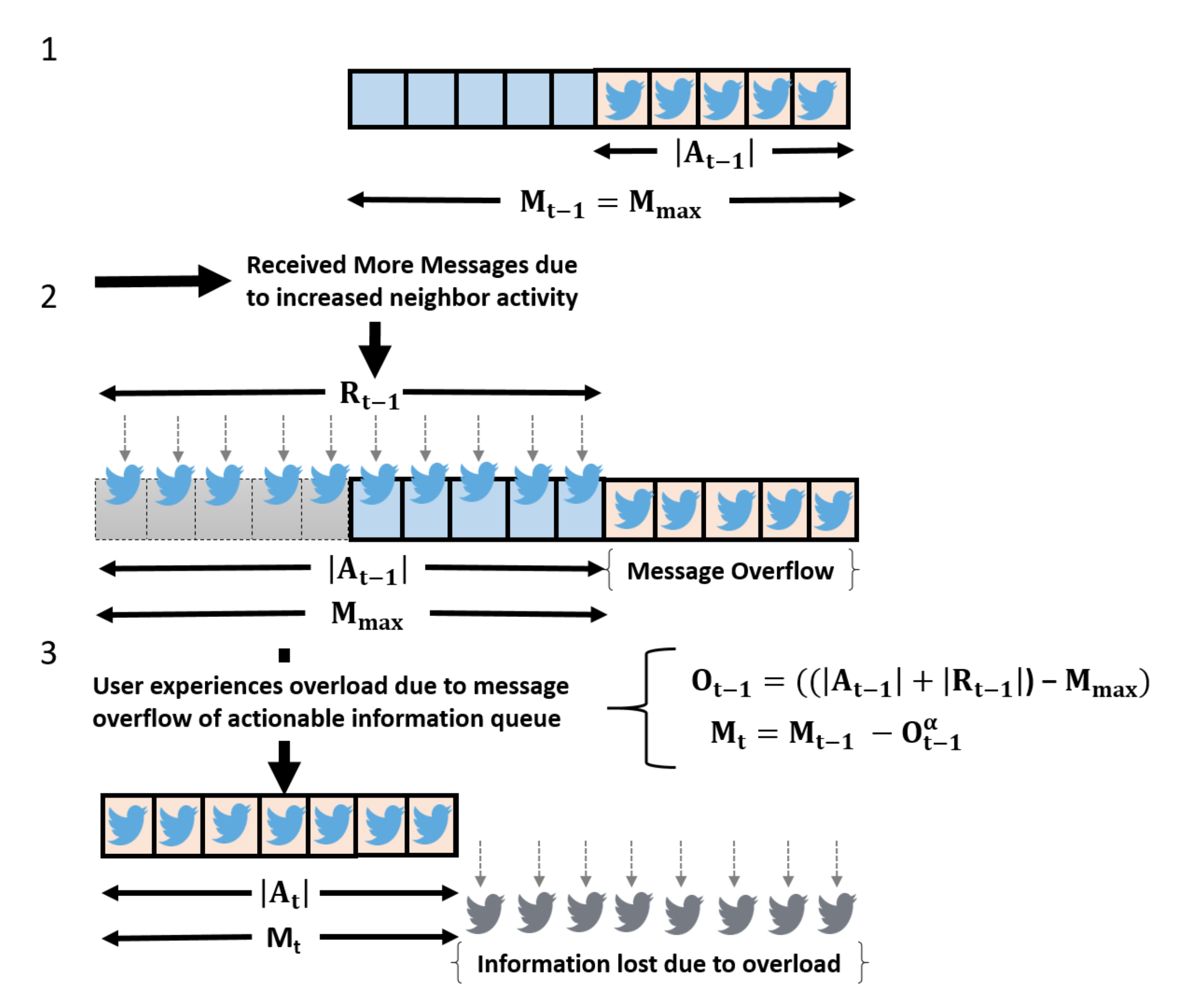}
    \caption{An illustrated demonstration of the actionable information queue and the process of information overloading. Step 1) Actionable information of capacity $M_t$ stores incoming information, and all messages are stochastically tried for response. If the individual is not overloaded $M_t = M_{max}$, where $M_{max}$ is the information overload threshold. 2) Received information is added to the front of the actionable information queue. A user is overloaded if $M_t$ is exceeded by the amount of actionable and received information, $|A_{t-1}| + |R_{t-1}|$, in which case a new $M_t$ is calculated based on the intensity of overload experienced, $O_{t-1}$, and rate of information loss, $\alpha$. 3) Excess messages are dropped in a first-in-first-out fashion, removing the oldest messages first.}
    \label{fig:infoprocess}
\end{figure*}

Fig. \ref{fig:infoprocess} summarizes the proposed mechanisms of information overload. Individuals modeled in the MACM often receive more than a single message every time step from multiple influencing neighbors. Often, the influence probability of the sender is insufficient for the users to act on the particular piece of information within the same time step it was received. Accordingly, all MACM agents have an actionable information queue which buffers incoming messages until they are processed and removed. At a give time step, each message in the queue is tried for response at the conditional probability of influence calculated by the MACM as described above. If new information is received by an agent while the actionable information queue is full, then the oldest messages are pushed out of the queue in a first-in-first-out manner to make room for the new, incoming messages.

In other words, when applied to a social media context, the current actionable information capacity at a given time $t$, $M_t$, represents the collection of incoming messages that a conversation participant may instinctively respond to, without actively searching archived messages in their notification feed. In order to simulate information overload, the following mechanism was utilized when determining $M_t$. At every time step, $M_t$ of each individual is re-calculated. First, the amount of excessive incoming information, $O_{t-1}$ experienced by the individual is quantified as shown in eq. \ref{eq:2}. An individual is considered to be overloaded if the total messages to consider, i.e. the sum of number of messages that were received from the previous time step ($R_{t-1}$) and the number of messages left over on the actionable information queue from the previous time step ($|A_{t-1}|$), exceed the information overload threshold, $M_{max}$. 
For overloaded individuals, $O_{t-1}^\alpha$  ($0 <= \alpha <= 1$) is defined as the amount of information lost due to an excess of incoming information past $M_{max}$, during $t$. A parameter of the model, $\alpha$ captures the power-law exponent found in \citep{gomez2014quantifying}, which, for overloaded individuals, represents the rate at which the information missed increases with an increasing excesses of incoming messages.
For un-overloaded individuals, $O_{t-1}$ is 0.
$O_{t-1}$ has a lower bound of 0 and upper bound of $M_{max}$. Gomez-Rodriguez et al. estimate the value of $M_{max}$ to be 30.

\begin{equation}
\label{eq:2}
O_{t-1} = \left\{
\begin{array}{@{}ll@{}}
    (|A_{t-1}| + |R_{t-1}|) - M_{max}, & \text{if}\ |A_{t-1}| + |R_{t-1}| >= M_{max} \\
     0, & \text{otherwise}
\end{array}\right.
\end{equation}

Next, the new actionable information capacity, $M_t$ is then calculated as shown in eq. \ref{eq:1}. If a user is overloaded (i.e. $O_{t-1} > 0$), then $M_t$ will be sub-optimal, that is $M_t < M_{max}$. $M_t$ has a lower bound of 0 and upper bound of $M_{max}$.

\begin{equation}
\label{eq:1}
M_{t} = \left\{
\begin{array}{@{}ll@{}}
    M_{t-1} - O_{t-1}^\alpha, & \text{if}\ O_{t-1}^\alpha <= M_{t-1} \\
    0, & \text{otherwise}
\end{array}\right.
\end{equation}

 
If $|A_{t-1}| + |R_{t-1}| > M_{t}$, i.e. there is an excessive number of messages to be considered for response, the oldest messages in the actionable information queue are removed until $|A_{t}|=M_{t}$. In other words, we assume that overloaded individuals prioritize responses based on recency, for which there is some supporting evidence in the literature \citep{leskovec2009meme}. In the following experiments, both $M_{max}$ and $\alpha$ are treated as parameters of the model, and the MACM along with the model of information overload are calibrated and analyzed in unison, as a single, ensemble model.

\section{Experiments}

We performed several experiments on the model ensemble described in the previous section. These experiments were conducted to: 1) provide a better understanding of the effects of the mechanisms of information overload over asynchronous communication media on online conversation dynamics, 2) identify possible values for $M_{max}$ and $\alpha$ in a real-world context, and 3) use the model along with the identified parameter values to reproduce and provide explanations to several phenomena observed in online conversations. In the following sections we use the terms \textit{conversations} and \textit{information cascades} interchangeably, considering that conversations are the means through which information cascades occur in human conversation. 

\label{sec:experiments}


\subsection{Data}
In order to achieve the above goals, we compared the output of this model to a real-world dataset of events pertaining to Twitter conversations among a cyptocurrency interest-community. This dataset consisted of a community of 1661 Twitter profiles engaged in discussions regarding the Electroneum cryptocurrency. Twitter conversations were between these users over a four month period starting from the 1\textsuperscript{st} of February, 2018 to the 1\textsuperscript{st} of June, 2018 were extracted. These conversations were used to extract the endogenous influence probabilities and construct the network for the MACM model as described in Sec. \ref{sec:1.1}. This data consisted of a total of 85083 events over 61380 unique conversations. Each event recorded an anonymized identifier of the user performing the event, the anonymized identifiers of the resulting conversation node, the parent being to replied, and the root tweet of the conversation, the type of action being performed by the user, and the timestamp. The trained model was then used to simulate a holdout period from the 1\textsuperscript{st} of June, 2018 to the 1\textsuperscript{st} of July, 2018 under varying parameter settings as described below. The time resolution used for simulations was 1 timestep equal to 1 hour. Actions of the events in this data were abstracted as follows: 1) Tweet events were considered as conversation initiation events, 2) Reply and quote events were considered as contributions to existing conversations, and 3) Retweet events were considered as conversation sharing actions. 



\subsection{Model Sensitivity}

In this section we discuss the sensitivity of conversation characteristics to $M_{max}$ and $\alpha$ and the relationships these model parameters have to user responsiveness.

\subsubsection{Sensitivity of Conversation Characteristics to $M_{max}$ and $\alpha$}
\label{sec:hypo_convo_charac}
We selected three measures characterizing conversation size, shape, and user participation for sensitivity analysis against the model parameters. In particular, the sensitivity of three information cascade properties, \textit{volume}, \textit{virality}, and the \textit{number of unique participants} to $M_{max}$ and $\alpha$, were tested.

The model was run for $\alpha$ varying over a range of  [0.0,0.9] with increments of 0.1 and $M_{max}$ varying over a range of [5,35] with increments of 5. Each parameter configuration was rerun for 10 repetitions. Since each run produced around 4000 unique conversations, approximately 40000 conversations were analyzed for each parameter configuration. The sensitivity of conversation volume, virality, unique participants, and responsiveness to $M_{max}$ and $\alpha$ were quantified using Sobol's sensitivity analysis technique \citep{sobol2001global}, and the first-order indices were compared to evaluate conversation sensitivity to the parameters. The volume of information cascades was measured as the count of contribution and sharing events appending information to the conversation created by an initiation event, inclusive of the initiation event itself. Cascade virality was measured with the Wiener index \citep{wiener1947structural,mohar1988compute,goel2015structural}, characterizing the shape of the tree structure formed by the conversation's appended contribution and sharing events. Measuring the number of unique users involved in an information cascade was simply done by counting the number of users engaged in a particular conversation. An individual's responsiveness was quantified as the probability that a received message would end up as the influencing parent of an outgoing message of an event performed by the individual, by tracking messages as they moved through user's actionable information queues. In order to understand the robustness and sensitivity of responsiveness to the model parameters, we measured the first order Sobol indices of $M_{max}$ and $\alpha$ as functions of one another, respectively.

\begin{figure}[h!]
    \centering
    \begin{subfigure}{0.5\textwidth}
        \centering
        \includegraphics[width=1\textwidth]{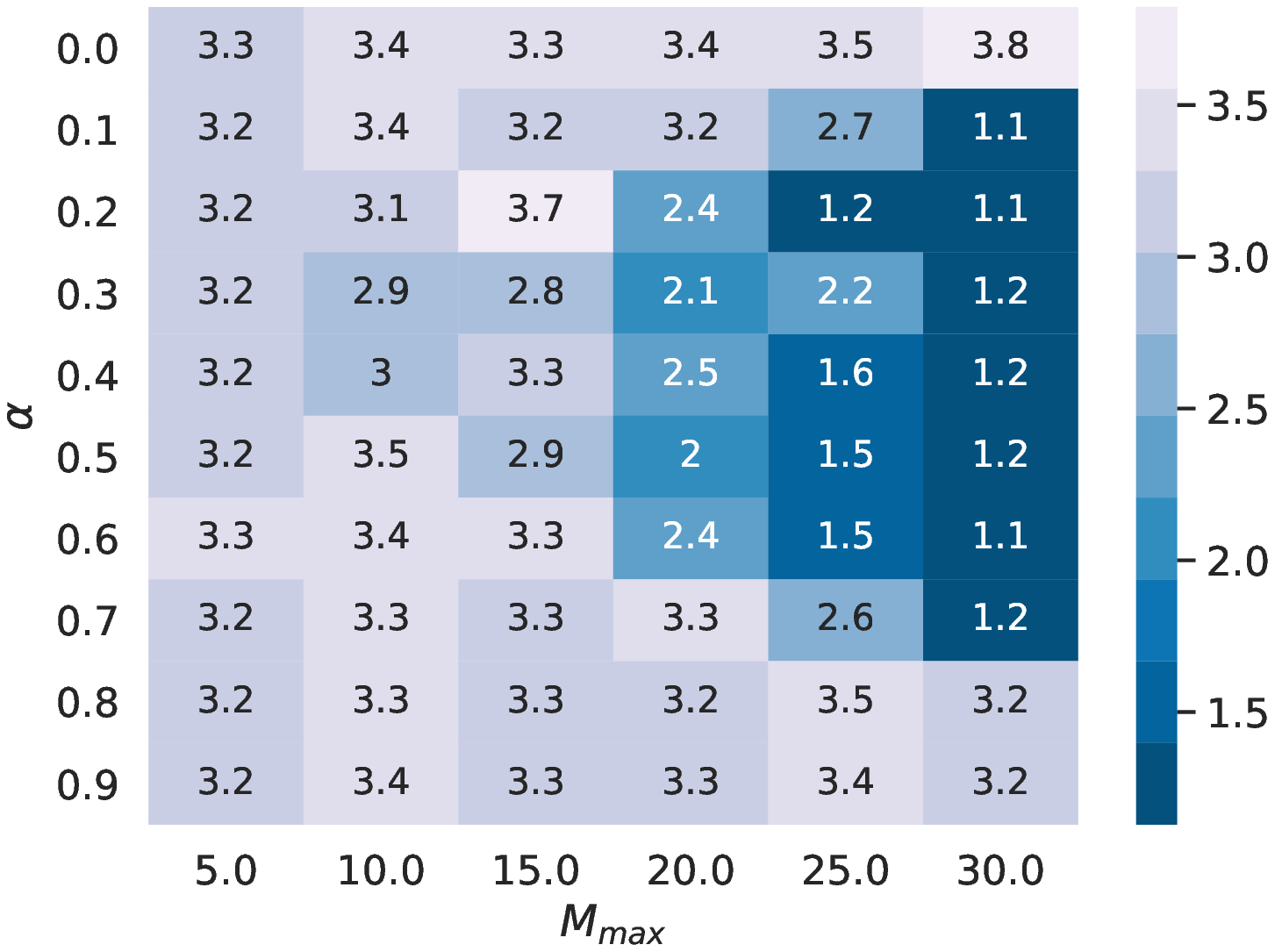}
        \caption{Simulated Twitter cascade volumes.}
    \end{subfigure}%
    ~~
    \begin{subfigure}{0.5\textwidth}
        \centering
        \includegraphics[width=1\textwidth]{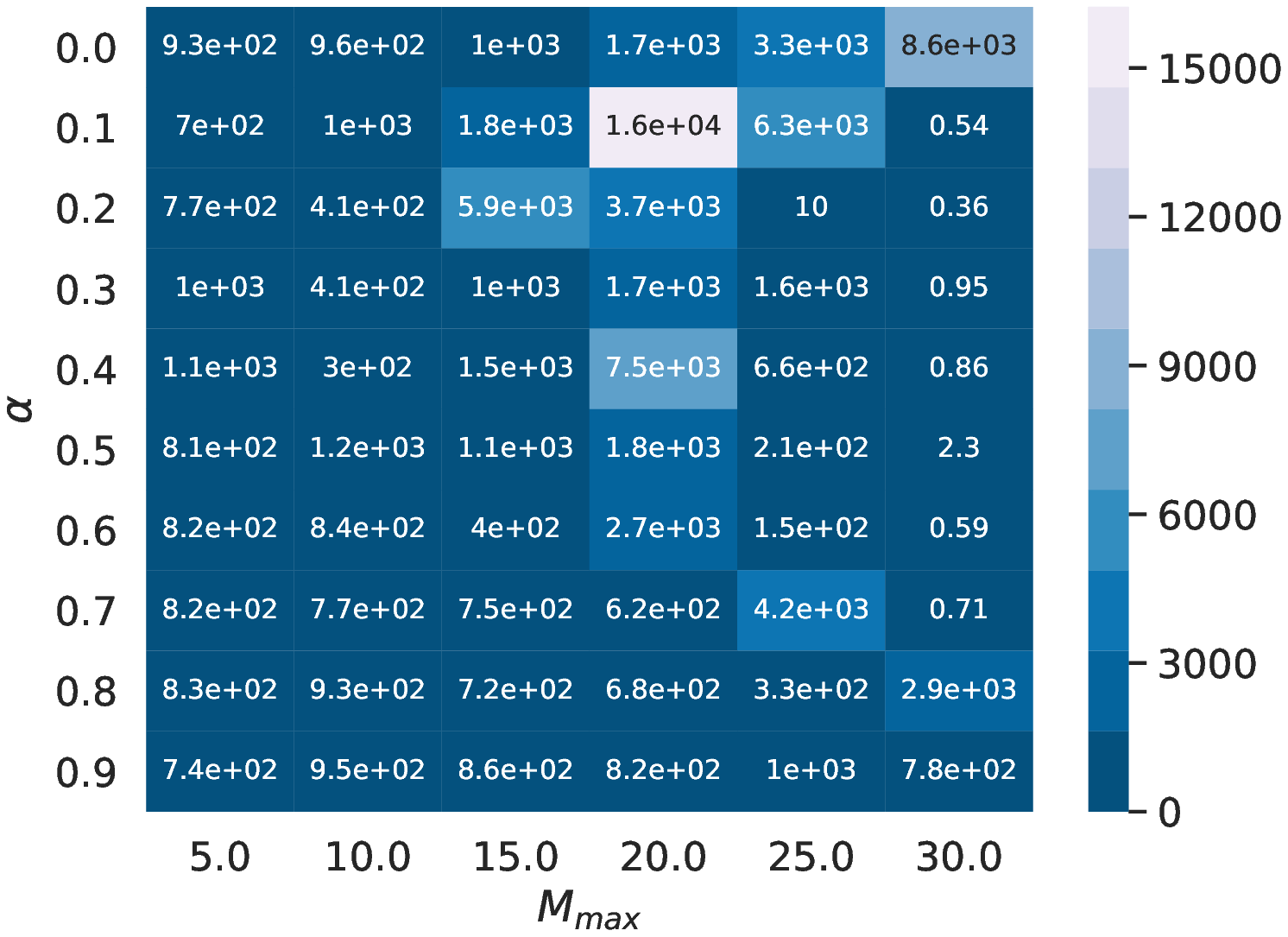}
        \caption{Simulated Twitter cascade virality (Wiener index).}
    \end{subfigure}%
    
    \begin{subfigure}{0.5\textwidth}
        \centering
        \includegraphics[width=1\textwidth]{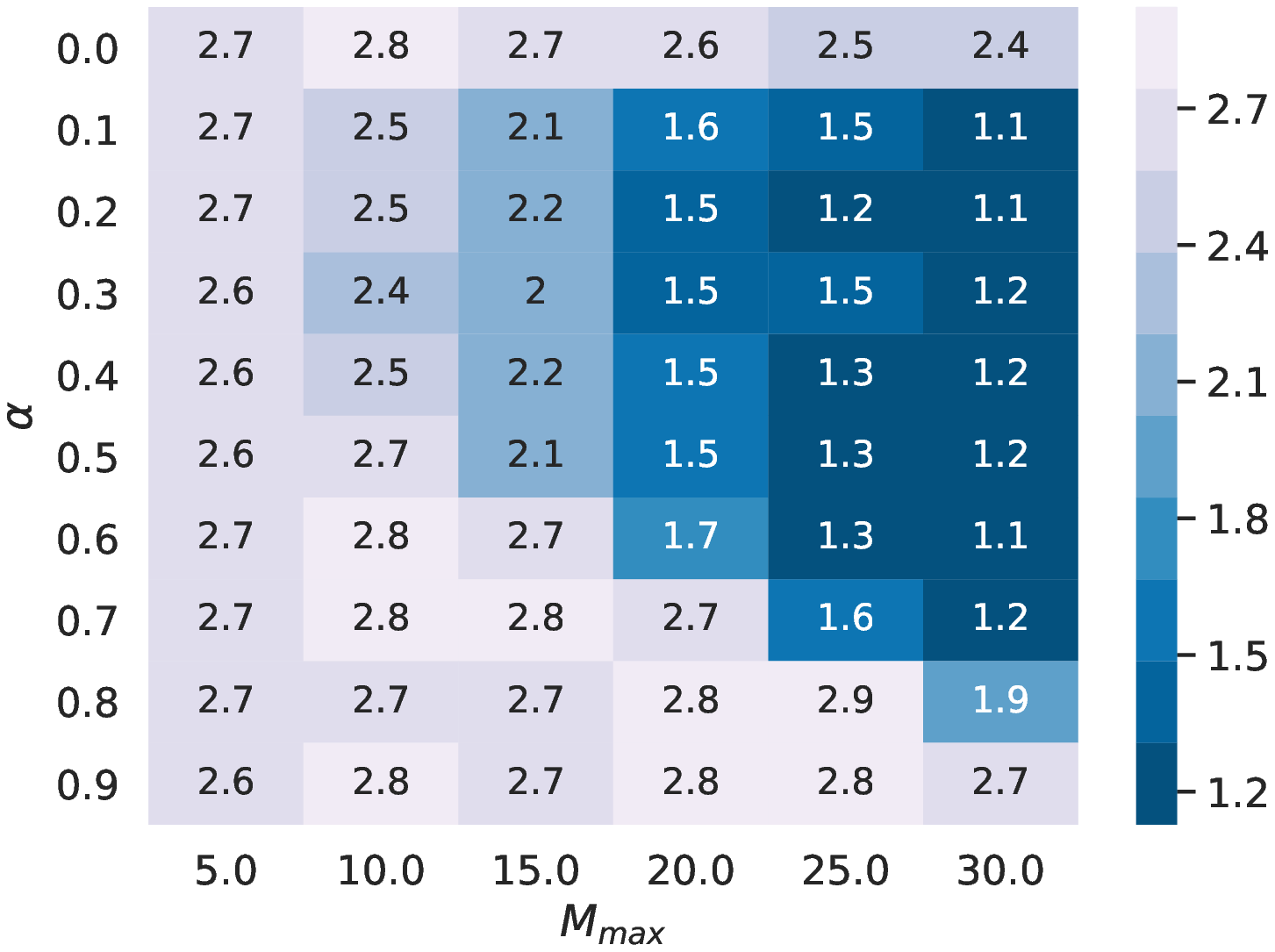}
        \caption{Unique participant counts on simulated Twitter cascades.}
    \end{subfigure}%
    \caption{Cascade volume, virality, and unique user counts over varying $M_{max}$ and $\alpha$. Cascade volume and unique user counts show sensitivity, with lower values for higher $M_{max}$ and mid-range $\alpha$, whereas cascade virality shows no relationship.}
    \label{fig:cascademeasures}
\end{figure}

Fig. \ref{fig:cascademeasures} displays the cascade volume, virality, and unique user counts over varying $M_{max}$ and $\alpha$ values. A strong relationship is seen between the parameters and unique user counts, while a weaker relationship is seen between the parameters and cascade volume. No relationships is seen between cascade virality and either parameter. Higher $M_{max}$ and mid-range $\alpha$ are shown to produce smaller cascades, i.e. shorter conversations, between less users.  

Tab. \ref{tab:1} displays the first order ($S_i$) and total ($S_{Ti}$) Sobol sensitivity indices for $M_{max}$ and $\alpha$ on the means of each conversation characteristic. Overall, all three characteristics are more sensitive to $M_{max}$ than they are to $\alpha$. According to the $S_i$ values, it can be seen that mean unique users in cascades is the characteristic that is by far the most sensitive to first order changes in both $M_{max}$ and $\alpha$. The total of the first order indices $\sum{S_i}$ for both cascade volume and virality are much lower than 1, indicating that there is much uncertainty of both these cascade characteristics, regardless of either parameter. In contrast, $\sum{S_i}$ is quite high for the mean number of unique users, indicating that this cascade characteristic has little uncertainty for varying values of $M_{max}$ and $\alpha$. Finally, $\sum{S_{Ti}}$ is greater than 1 for all cascade characteristics indicating that there is some level of sensitivity to the interaction of both $M_{max}$ and $\alpha$. 

\begin{table}[h!]
\centering
\caption{First order ($S_i$) and total ($S_{Ti}$) Sobol sensitivity indices for $M_{max}$ and $\alpha$ on mean cascade volume, mean cascade virality, and mean number of unique users participating in a cascade.}
\label{tab:1}       
\begin{tabular}{r| ll|ll|cc}
\hline\noalign{\smallskip}
    & \multicolumn{2}{c|}{$M_{max}$}& \multicolumn{2}{c|}{$\alpha$}&\multicolumn{2}{c}{Total}  \\
\noalign{\smallskip}\hline\noalign{\smallskip}
    & $S_i$& $S_{Ti}$ & $S_i$& $S_{Ti}$ & $\sum{S_i}$&$\sum{S_{Ti}}$ \\
\noalign{\smallskip}\hline\noalign{\smallskip}
Mean Cascade Volume & 0.035 & 0.986 & 0.014 & 0.965 & 0.049 & 1.951\\
Mean Cascade Virality & 0.074 & 0.980 & 0.020 & 0.926 & 0.094 & 1.906\\
Mean Unique Users in Cascade & 0.642 & 0.769 & 0.231& 0.358 & 0.873 & 1.127\\
\noalign{\smallskip}\hline
\end{tabular}
\end{table}


\subsubsection{Relationship of Responsiveness to $M_{max}$ and $\alpha$}
\label{sec:hypo_resp}
The theoretical proof by Li et al. \citep{li2014modeling} states that the probability of responding to a neighbor's incoming message, or \textit{responsiveness} as defined in \citep{gomez2014quantifying}, is unrelated to the size of `view scope', represented here by $M_{max}$, under information overload. However, Li et al., implicitly assume that there is no loss of actionable information due to information overload. We argue that this only represents the case of $\alpha = 0$. In the model described above, under the condition $\alpha > 0$, the information missed by an individual would increase by the amount of excessive information raised to the power $\alpha$, causing the probability of response to any particular message to drop. Therefore, a larger $M_{max}$ would provide more robustness against higher $\alpha$.

\begin{figure*}[h!]
\centering
    \begin{subfigure}{0.5\textwidth}
        \centering
        \includegraphics[width=\textwidth]{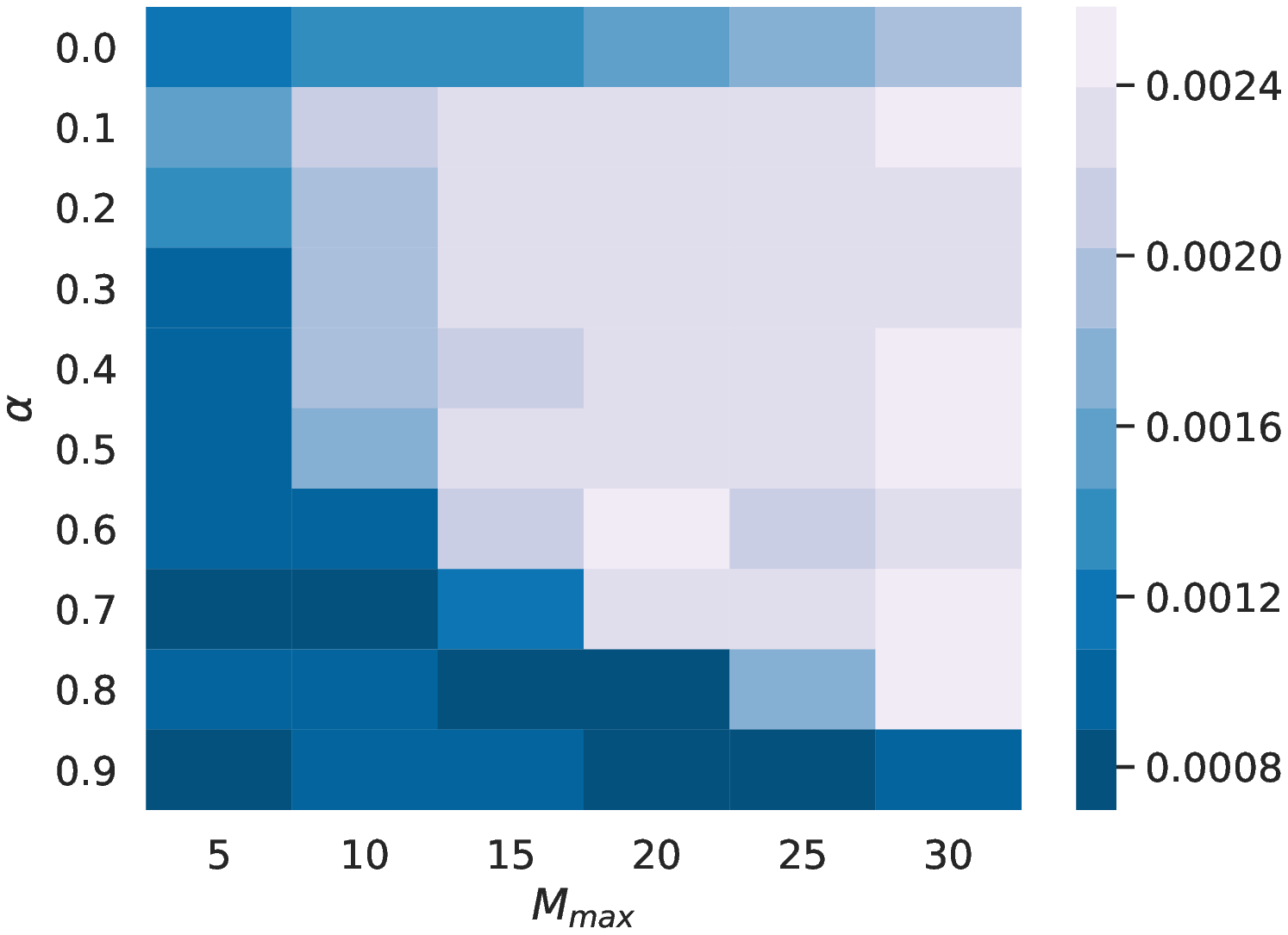}
        \caption{Mean responsiveness of individuals under varying $M_{max}$ and $\alpha$.}
        \label{fig:response_prob_mean}
    \end{subfigure}%
    ~~
    \begin{subfigure}{0.5\textwidth}
        \centering
        \includegraphics[width=\textwidth]{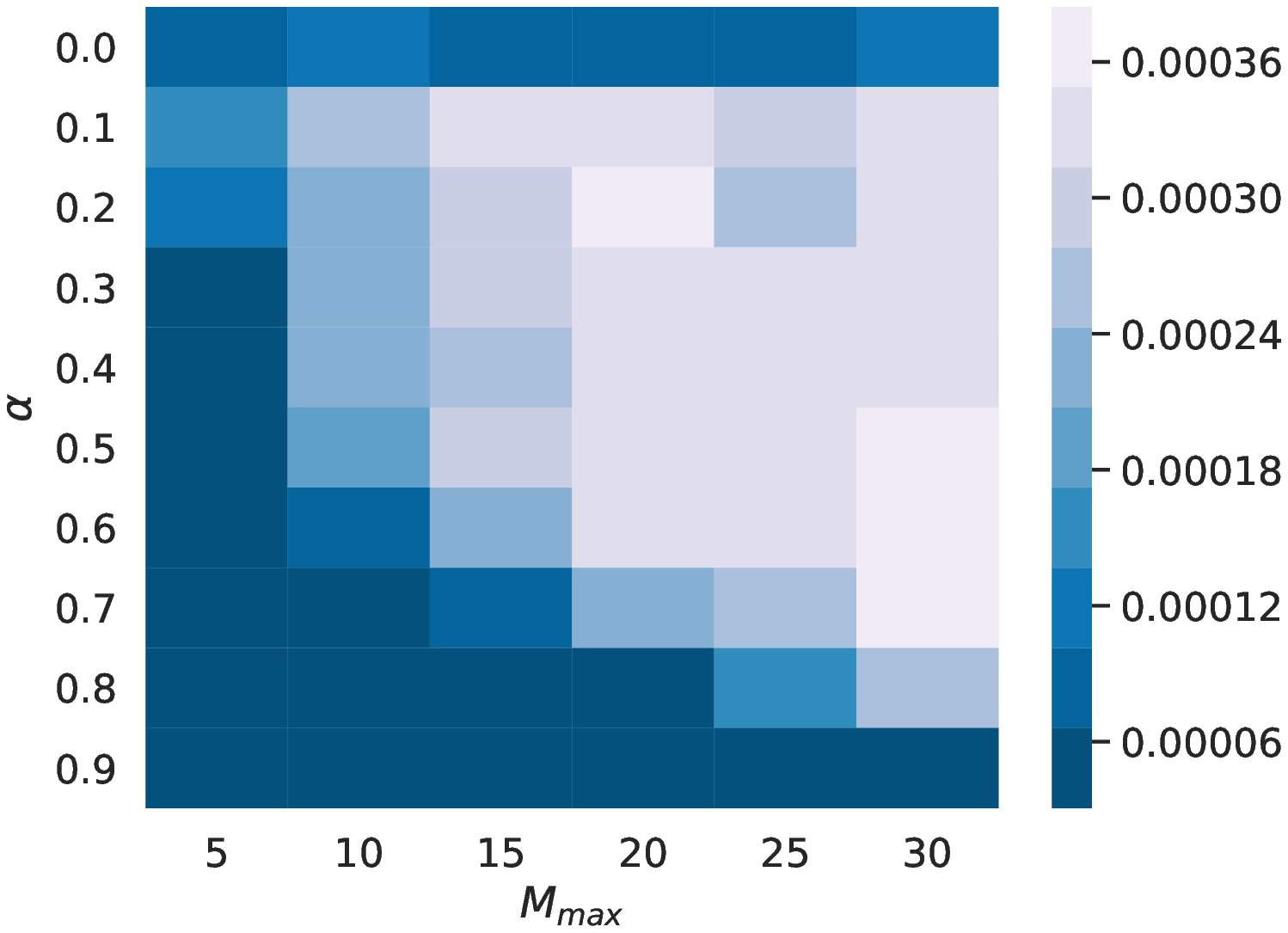}
        \caption{Variance of responsiveness of individuals under varying $M_{max}$ and $\alpha$.}
        \label{fig:response_prob_var}
    \end{subfigure}
    \caption{Mean and variance of responsiveness of individuals under varying $M_{max}$ and $\alpha$. Responsiveness follows different relationships with $M_{max}$ when $\alpha=0$ and when $\alpha>0$. Under $\alpha>0$, higher values of $M_{max}$ provide higher robustness to responsiveness under overload.}
\end{figure*}

\begin{figure*}[h!]
\centering
    \begin{subfigure}{0.5\textwidth}
        \centering
        \includegraphics[width=\textwidth]{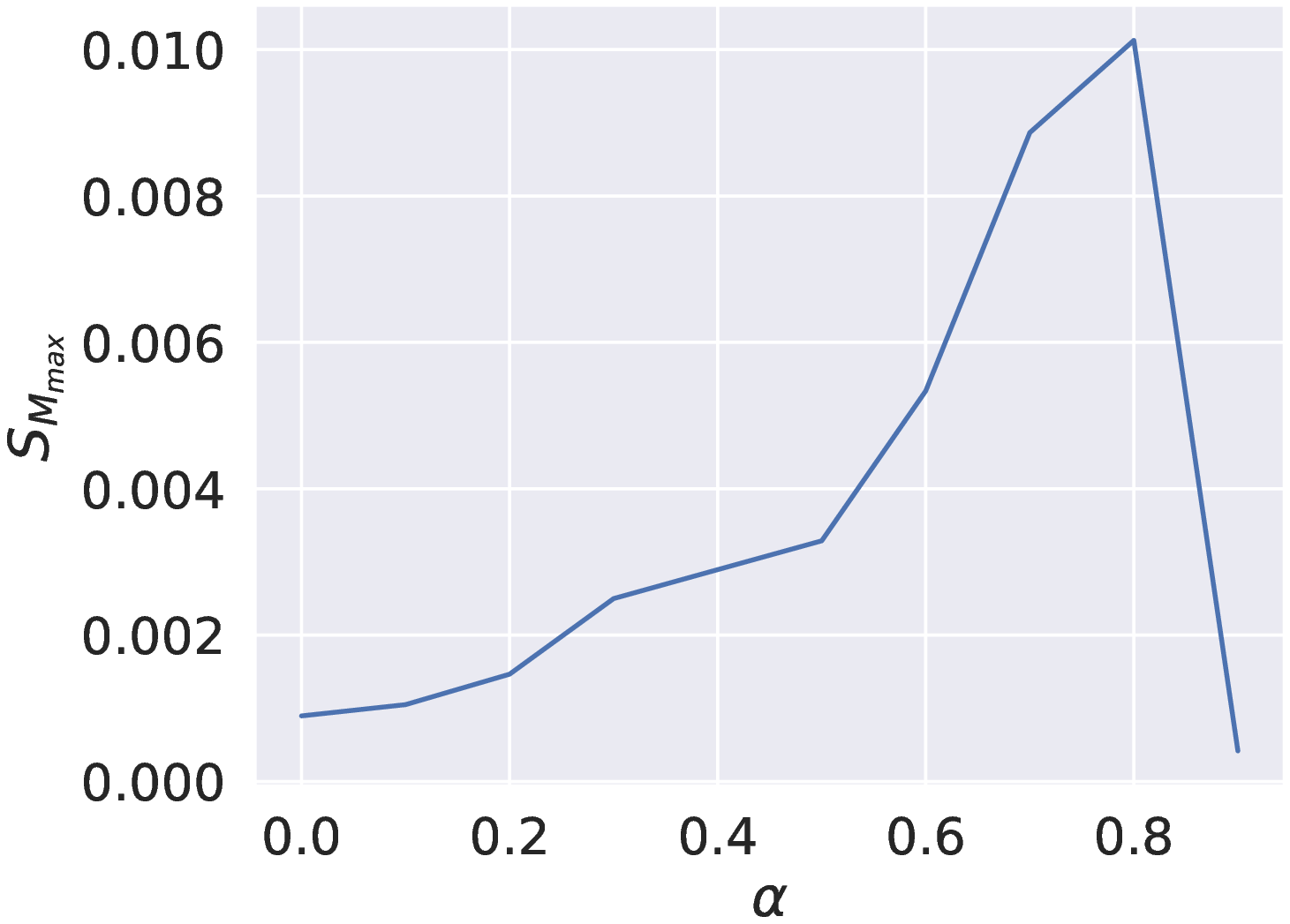}
        \caption{First order Sobol sensitivity index of responsiveness to $M_{max}$ under varying $\alpha$.}
        \label{fig:s1responsiveness_mt}
    \end{subfigure}%
    ~~
    \begin{subfigure}{0.5\textwidth}
        \centering
        \includegraphics[width=\textwidth]{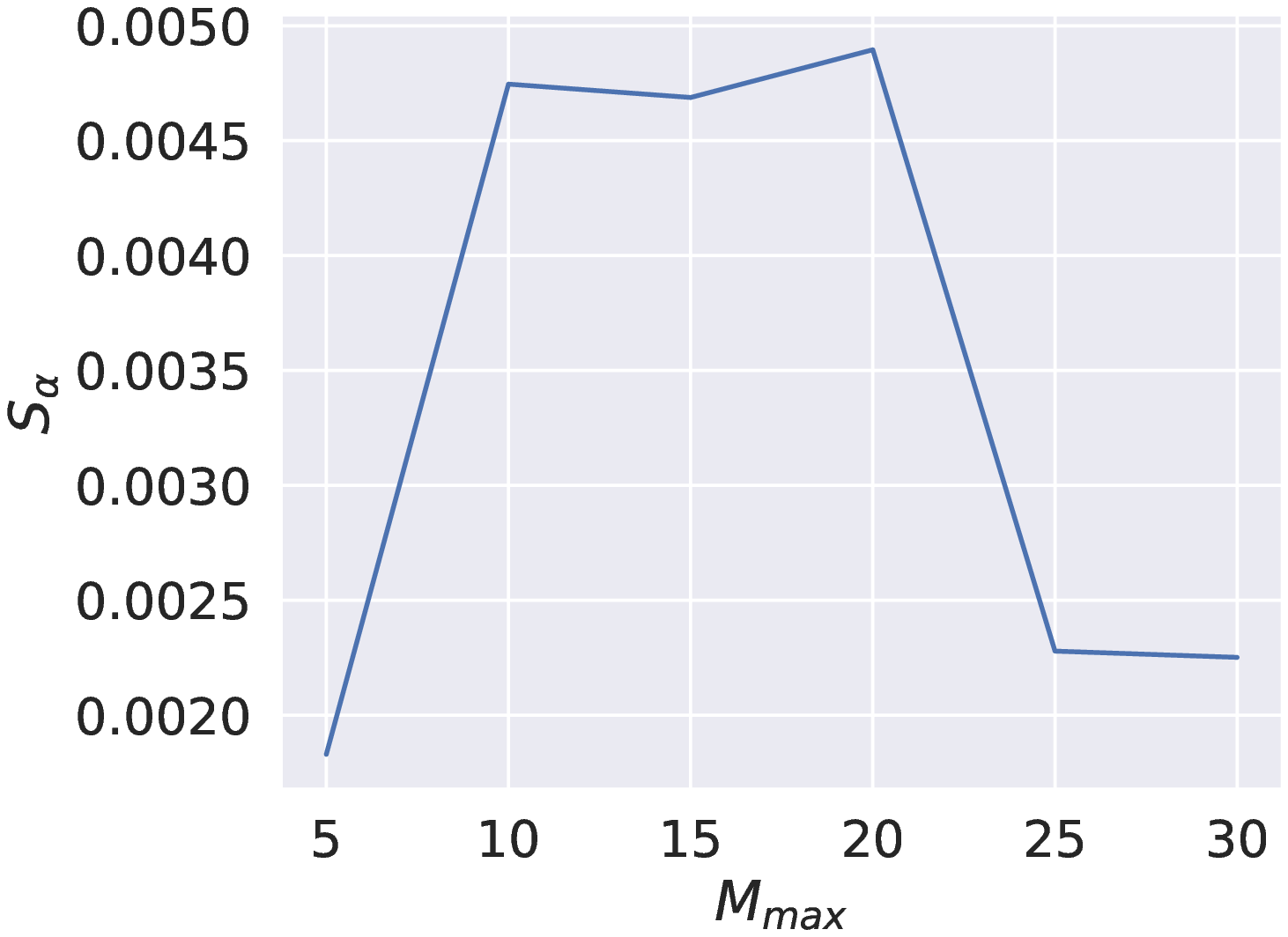}
        \caption{Variance of responsiveness of individuals under varying $M_{max}$ and $\alpha$.}
        \label{fig:s1responsiveness_alpha}
    \end{subfigure}
    \caption{First order Sobol sensitivity index of responsiveness to $M_{max}$ and $\alpha$. It is seen that sensitivity of responsiveness to $M_{max}$ increases non-linearly with increasing $\alpha$ until it peaks at 0.8, after which there is a rapid decline in sensitivity of responsiveness to $M_{max}$. In contrast, sensitivity of responsiveness to $\alpha$ is seen to be highest for moderate ranges of $M_{max}$, and robust under high $M_{max}$.  }
    \label{fig:s1responsiveness}
\end{figure*}

Fig. \ref{fig:response_prob_mean} displays the mean, and Fig. \ref{fig:response_prob_var} the variance, of the responsiveness of individuals to senders as $M_{max}$ and $\alpha$ are varied. It can be seen that responsiveness exists in two regimes at $\alpha=0$ and at $\alpha > 0$, as expected
. This indicates that when the assumption made in \citep{li2014modeling} is relaxed, i.e. the information overload threshold or view scope, $M_{max}$, \textit{is} suppressed under information overload, then responsiveness is in fact substantially sensitive to $M_{max}$. Accordingly, as shown in Fig. \ref{fig:s1responsiveness_mt}, the first order Sobol sensitivity indices for $M_{max}$ increase non-linearly as $\alpha$ increases until $\alpha=0.8$, 
after which it drops drastically. This indicates that the sensitivity of responsiveness to capacity increases non-linearly with the rate of information loss up to a threshold. Beyond this threshold, the rate of loss of information is so harsh, that even a minimal quantity of information inflow can completely overwhelm the user. Fig. \ref{fig:s1responsiveness_alpha} shows that responsiveness is fairly sensitive to $\alpha$ for mid range $M_{max}$. But when $M_{max}$ is substantially high, responsiveness is quite robust against $\alpha$. 
In other words, mid-range capacity allows for a fair amount of sensitivity to rate of information loss. But under high capacity, the effect of rate of information loss on responsiveness is diminished.



\subsection{Calibration and Analysis of Cryptocurrency Interest Community on Twitter}
\label{sec:hypo_calib}
The model was calibrated to find the best $M_{max}$ and $\alpha$ values that most closely mimicked conversation dynamics of the Electroneum interest community. This was done by performing a factorial grid experiment over the same parameter ranges defined above, but also calculating the Jensen-Shannon divergence of each measurement from that of the respective measurements applied to the ground-truth data for June, 2018. Simulations of the calibrated model were then analyzed, including traces of the agents received information and actionable information queues.

\subsubsection{Capacity and Rate of Information Loss of Community}
Through the calibration of the model of information overload, we expected to see a similar information overload threshold as discovered in prior research. According to \citep{gomez2014quantifying}, the threshold of incoming tweets after which a Twitter user experiences information overload is 30 Tweets an hour.



\begin{figure*}[h!]
\centering
    \begin{subfigure}{0.5\textwidth}
        \centering
        \includegraphics[width=\textwidth]{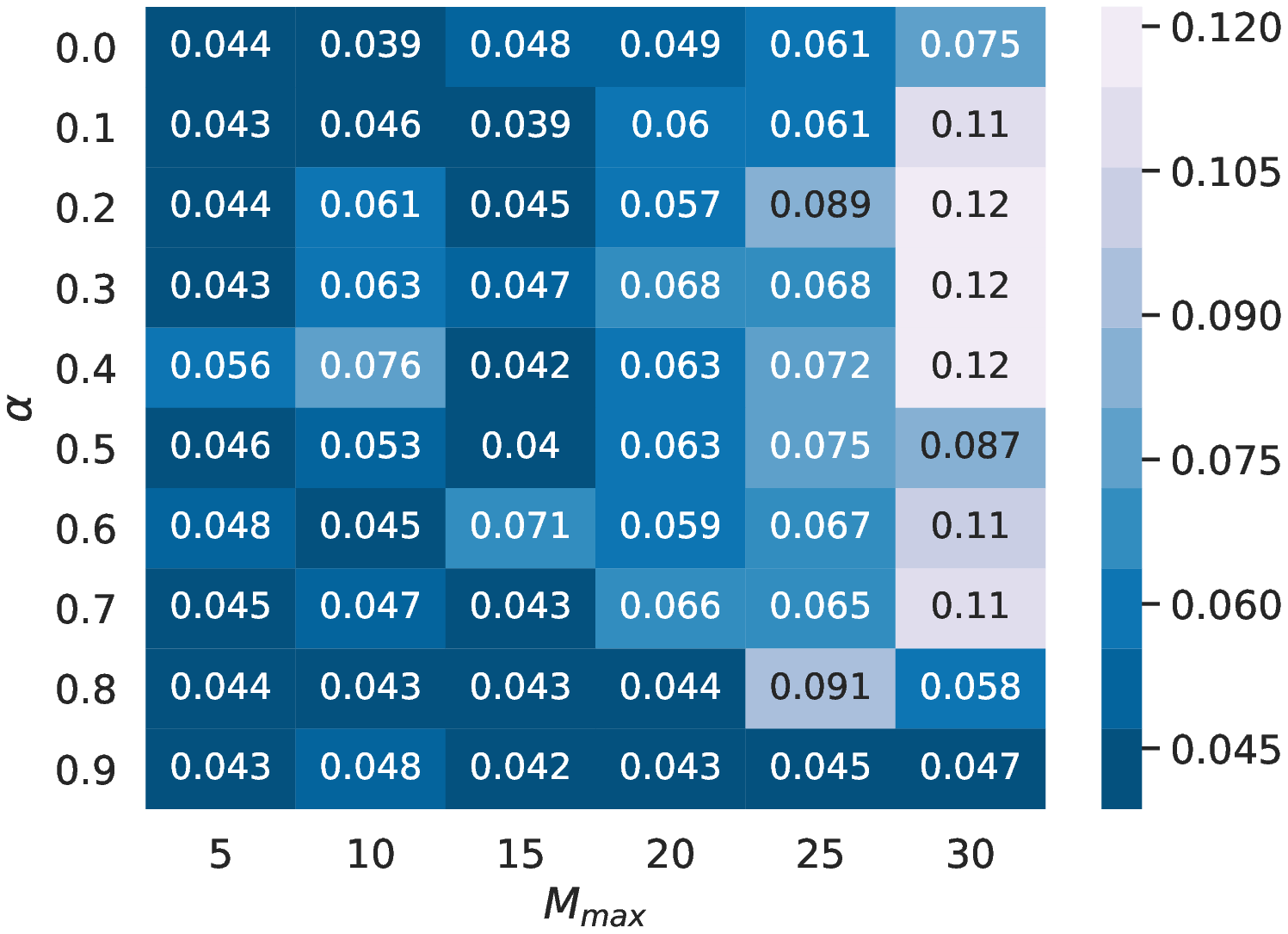}
        \caption{JS Divergence of simulated Twitter cascade volumes to the ground truth data, by $M_{max}$ and $\alpha$.}
    \end{subfigure}%
    ~~
    \begin{subfigure}{0.5\textwidth}
        \centering
        \includegraphics[width=\textwidth]{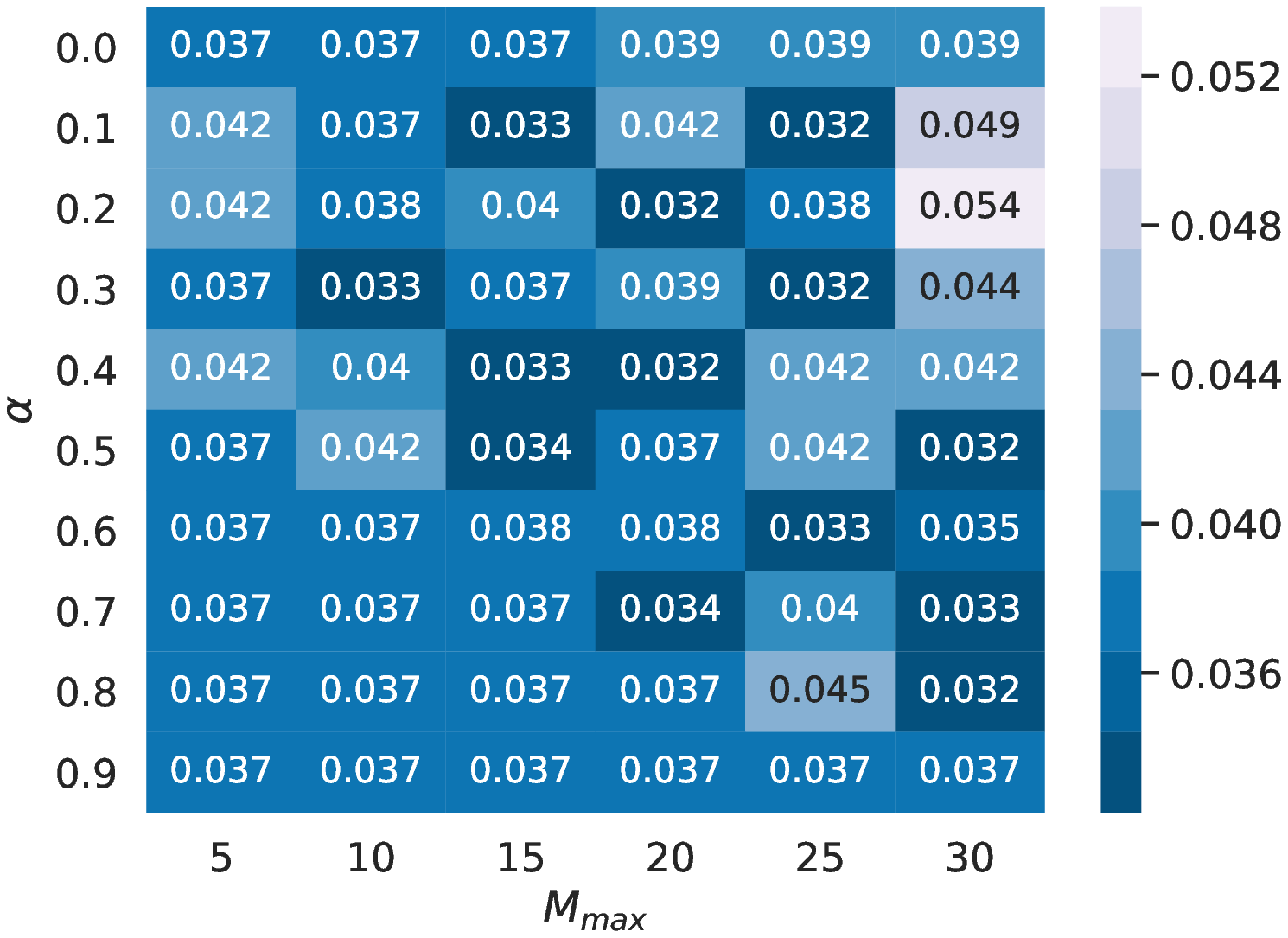}
        \caption{JS Divergence of simulated Twitter cascade virality values (Wiener index) to the ground truth data, by $M_{max}$ and $\alpha$.}
    \end{subfigure}
    
    \begin{subfigure}{0.49\textwidth}
        \centering
        \includegraphics[width=\textwidth]{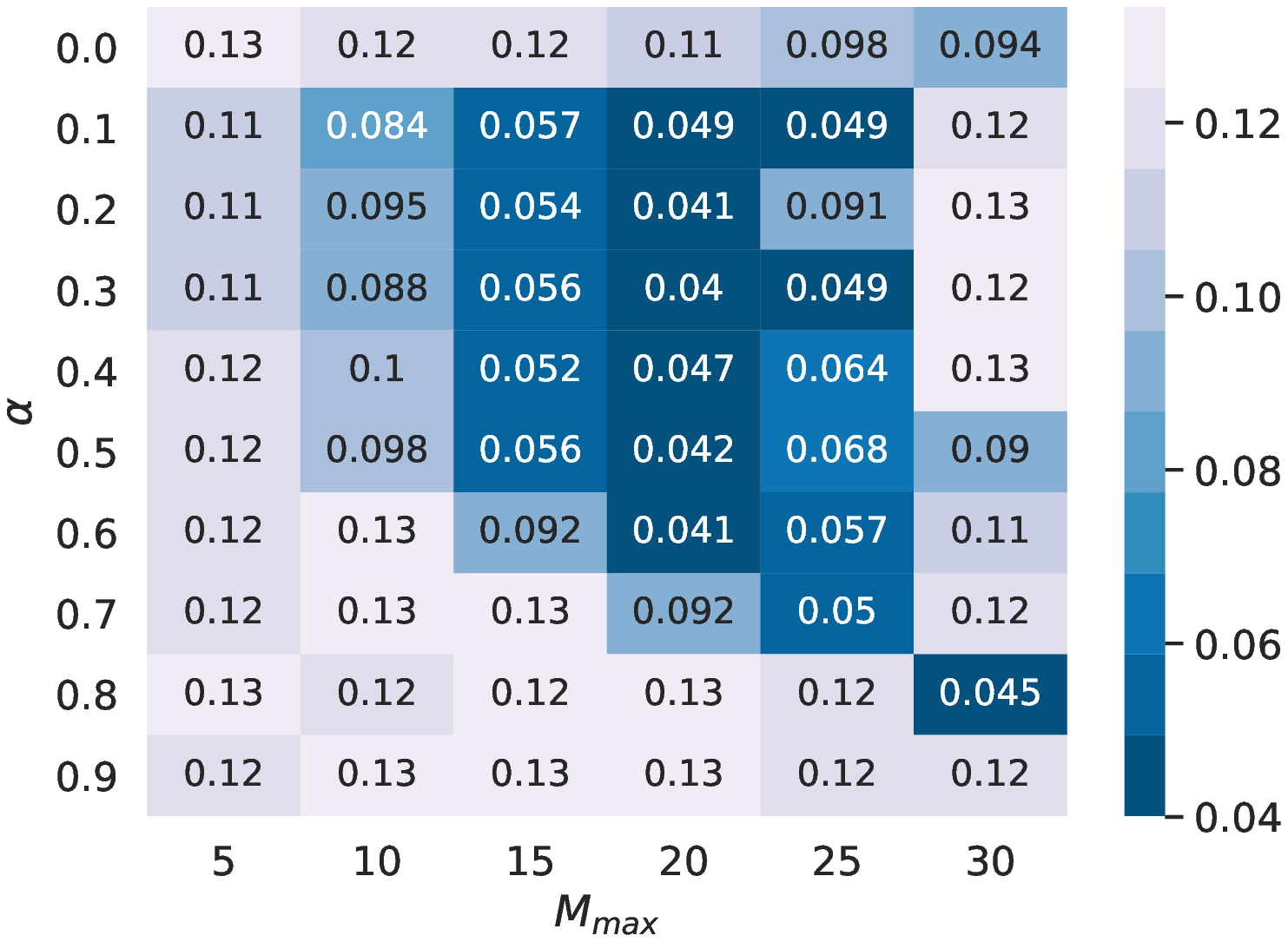}
        \caption{JS Divergence of unique user counts on simulated Twitter cascades to the ground truth data.\newline}
    \end{subfigure}
    \begin{subfigure}{0.49\textwidth}
        \centering
        \includegraphics[width=\textwidth]{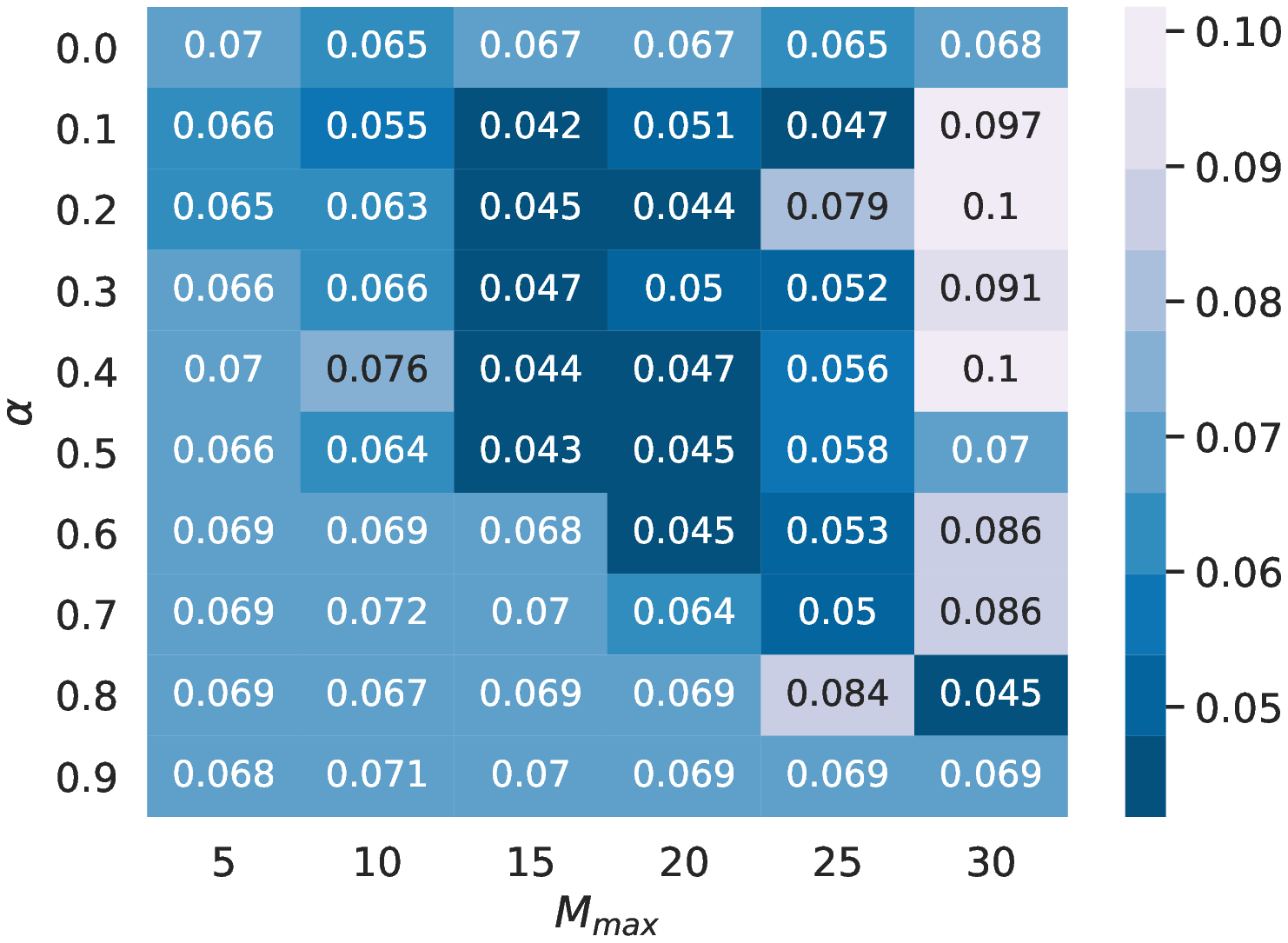}
        \caption{Mean of JS Divergence of cascade volumes, cascade virality, and unique user counts on simulated Twitter cascades to the ground truth data.}
    \end{subfigure}
    \caption{The JS divergence of simulated cascade volume, virality, and unique user counts against that of the ground truth data, over varying $M_{max}$ and $\alpha$. The optimal values of the JS divergence of unique user counts shows a Pareto-front between the two parameters. The lowest aggregate JS divergence is obtained at $M_{max}=30$, $\alpha=0.8$, which agrees with \citep{gomez2014quantifying} that Twitter user responsiveness is constant upto a threshold of 30 incoming tweets, after which the user appears to be overloaded.}
    \label{fig:cascademetrics}
\end{figure*}

The results of the factorial experiment on the JS divergence (lower is better) between the simulation and the ground truth measurements are shown in Fig. \ref{fig:cascademetrics}. There is a clear, defining relationship of the JS divergence of unique user counts on cascades and the two parameters. Together, $M_{max}$ and $\alpha$ form a Pareto-front, where highly fit simulations exist under a trade-off between the two parameters, from mid-range $M_{max}$ and low $\alpha$ to high $M_{max}$ and high $\alpha$. The relationships of both $M_{max}$ and $\alpha$ to JS divergence in cascade volume are generally constant until $M_{max}$ is increased past mid-range, under which JS divergence in volume is lowest for high $\alpha$. No apparent trend in JS divergence in conversation virality was seen. Most interestingly, 
when the measurements for each configuration are aggregated, by taking their means, the lowest JS divergence is reported for $M_{max}=30$, $\alpha=0.8$, which matches the information overload threshold for Twitter, 30 Tweets per hour, measured by Gomez-Rodriguez et al..

\begin{figure*}[h!]
\centering
  \includegraphics[width=0.5\textwidth]{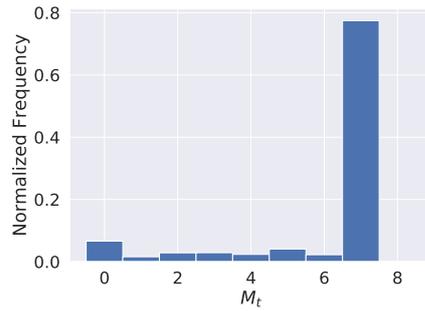}
\caption{Overall distribution of hourly actionable information queue capacity ($M_t$) for Twitter notifications of participants engaged in Electroneum discussions across the month of June, 2018. A maximal number of 7 is seen through the simulations, while relatively more overloaded states are uncommon, yet observed at which $M_t < 7$.}
\label{fig:extendedmemorycapacityhist}       
\end{figure*}

Furthermore, we find that there exists an upper limit in $M_t$ of 7 tweets per hour for simulations under the calibrated parameter values, as shown in Fig. \ref{fig:extendedmemorycapacityhist}. The overall distribution of $M_t$ is highly skewed towards this maximum value, with a few instances where users experienced stronger information overload and reported $M_t$ values of less than 7.

\subsubsection{Overloaded Individuals as Conveyors of Important Information}
\label{sec:hypo_overload_cascades}
In \citep{hodas2013friendship}, it is found that overloaded users are not good signalers for important information, as they receive more information from large, popular cascades than small, growing cascades. We expect the described ensemble model to demonstrate the same behavior, once calibrated to the real-world data. 


The explanatory nature of the agent-based model allowed us to further verify the cause of the above phenomena observed in \citep{hodas2013friendship} through analysis of overloaded users' actionable information queues. As the MACM simulated the information contained within the actionable information queues of all agents, we were able to analyze the agent's `memory' over time, a task not possible with lab experimentation. Moreover, we compared the volume, virality and number of unique participants of cascades contained within agents' actionable information queue against the rate of information loss under overload.

\begin{figure*}[h!]
\centering
    \begin{subfigure}{0.5\textwidth}
        \centering
        \includegraphics[width=\textwidth]{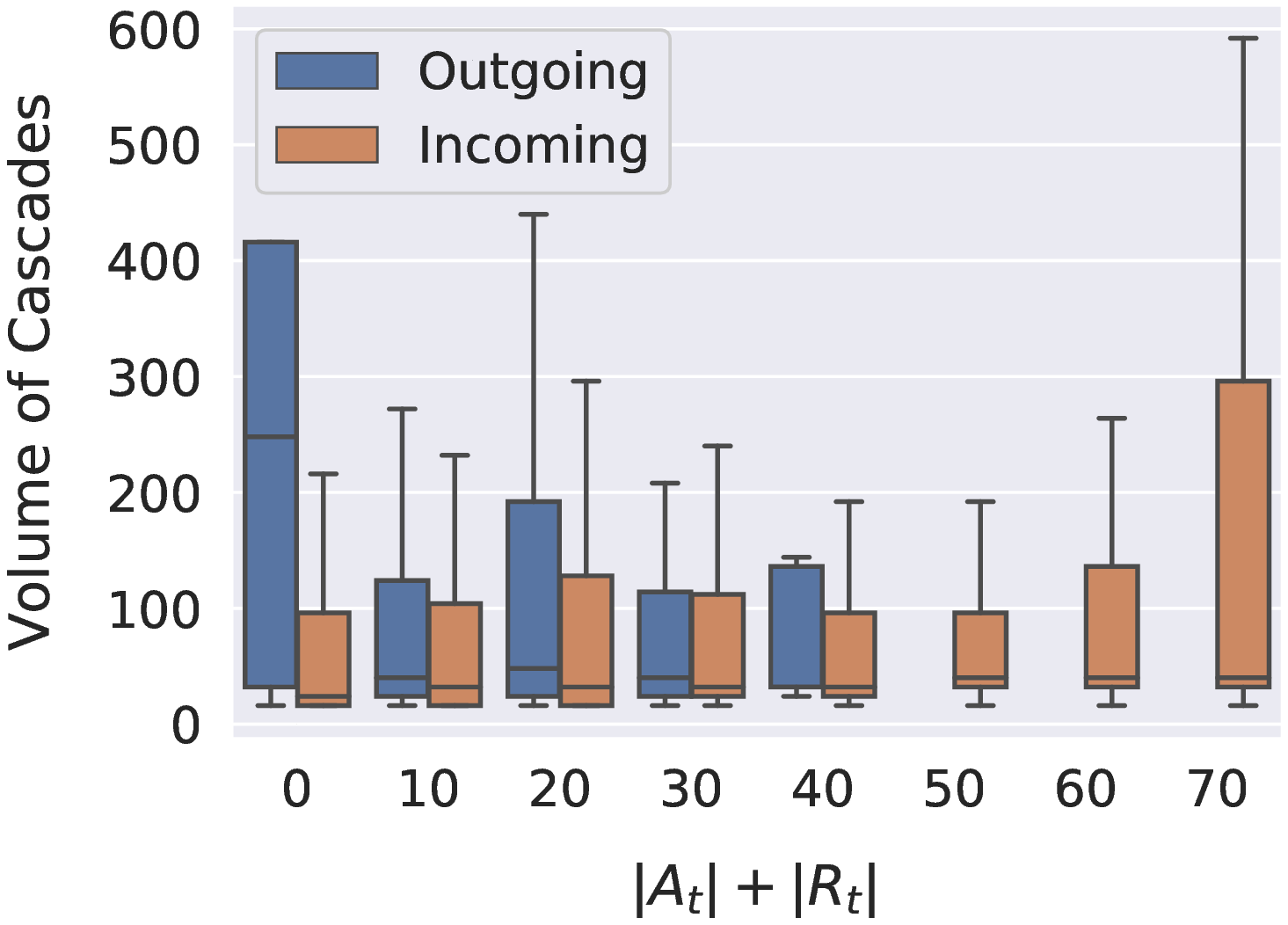}
        \caption{Cascade volume (outgoing in blue, incoming in orange) vs received and actionable information quantity.}
        \label{fig:vol_in_out_ai_ri}
    \end{subfigure}%
    ~~
    \begin{subfigure}{0.5\textwidth}
        \centering
        \includegraphics[width=\textwidth]{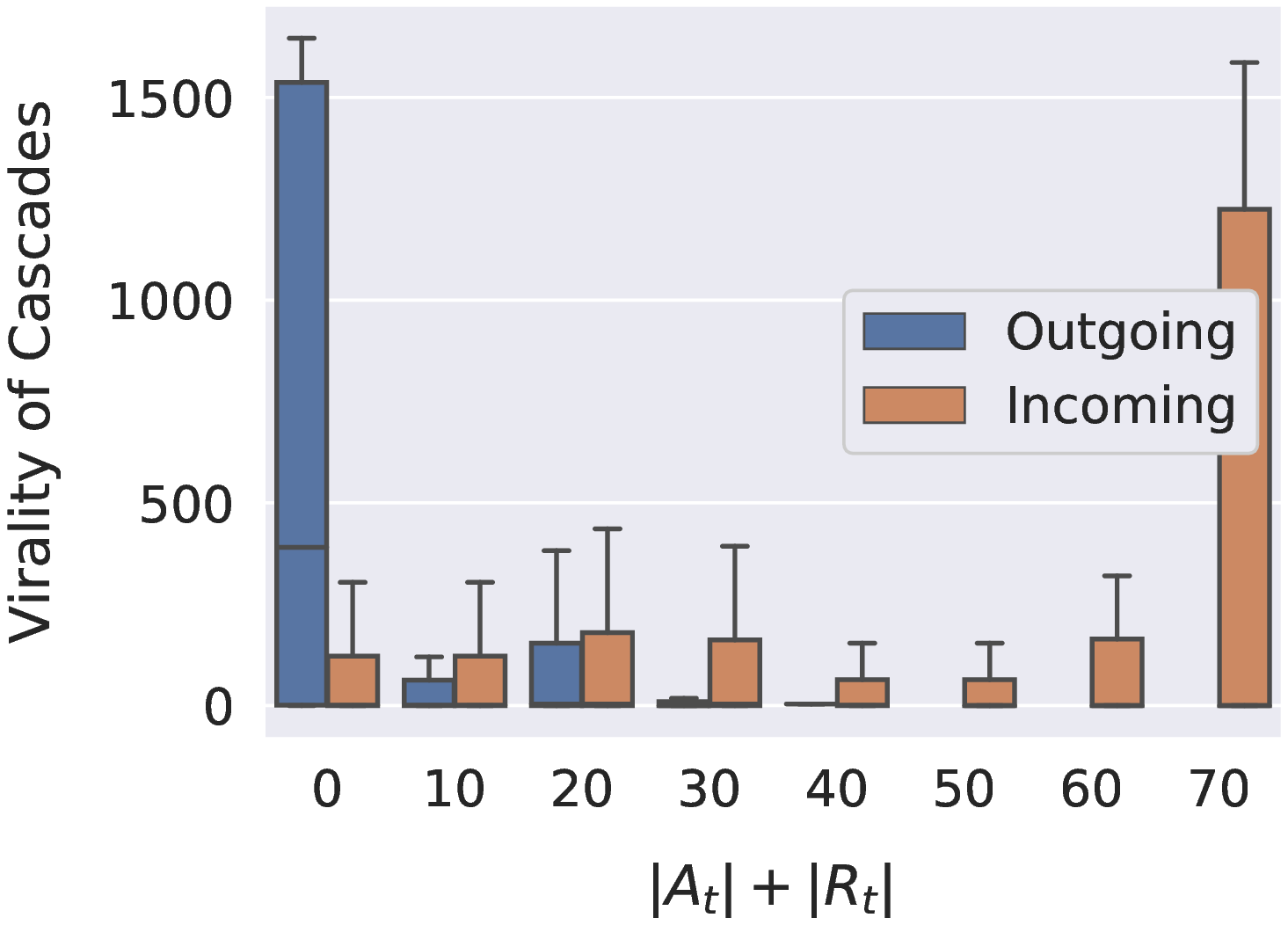}
        \caption{Cascade virality (outgoing in blue, incoming in orange) vs received and actionable information quantity.}
        \label{fig:vir_in_out_ai_ri}
    \end{subfigure}
    
    \begin{subfigure}{0.5\textwidth}
        \centering
        \includegraphics[width=\textwidth]{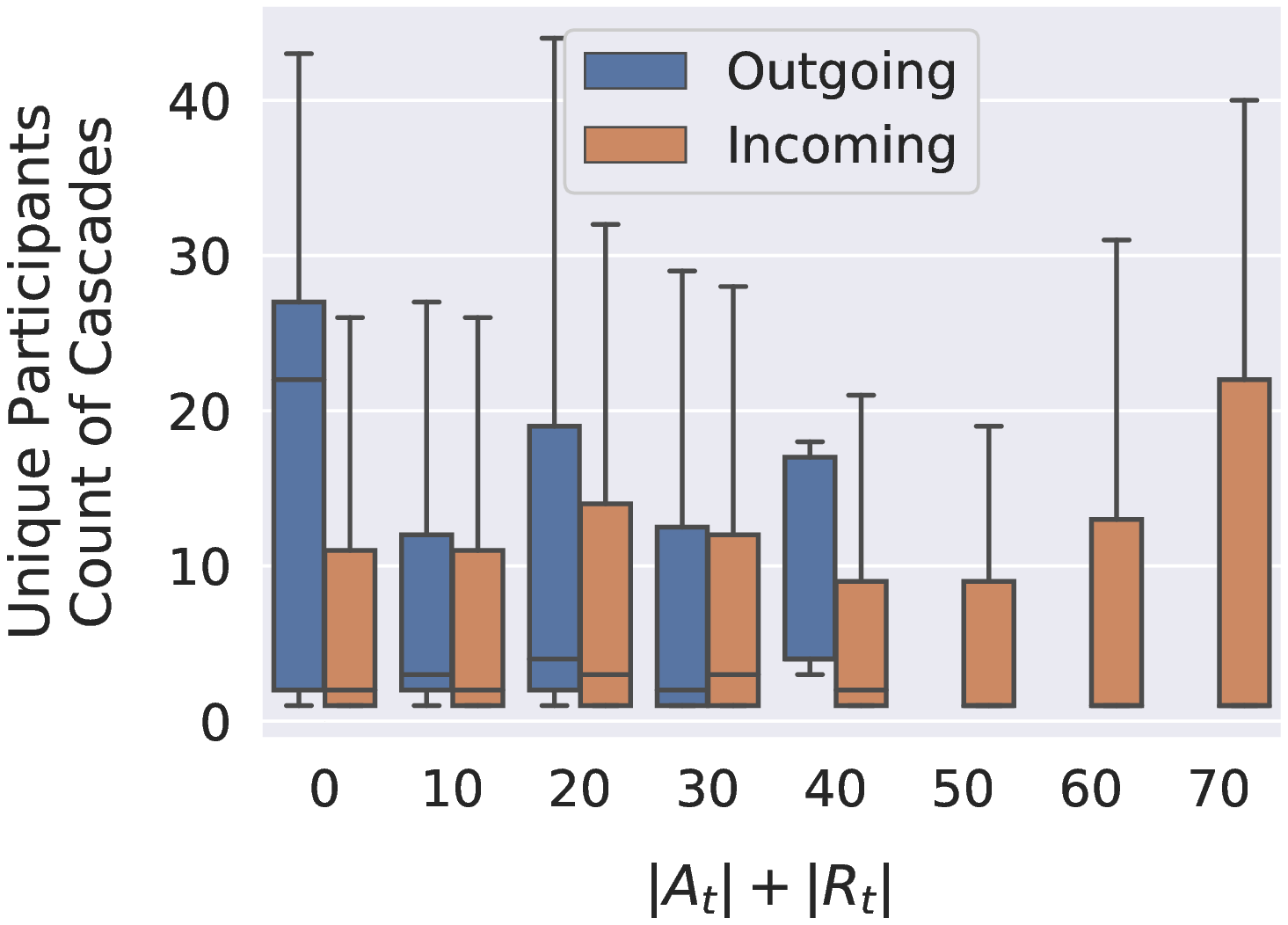}
        \caption{Unique cascade participants (outgoing in blue, incoming in orange) vs received and actionable information quantity.}
        \label{fig:uuc_in_out_ai_ri}
    \end{subfigure}
    \caption{Cascade characteristics against received and actionable information quantity. Users who are overloaded and receiving more information, received larger, popular, and more viral cascades but are unable to participate in them at the same rate.}
    \label{fig:volviruuc_in_out_ai_ri}
\end{figure*}

\begin{figure*}[h!]
\centering
    \begin{subfigure}{0.5\textwidth}
        \centering
        \includegraphics[width=\textwidth]{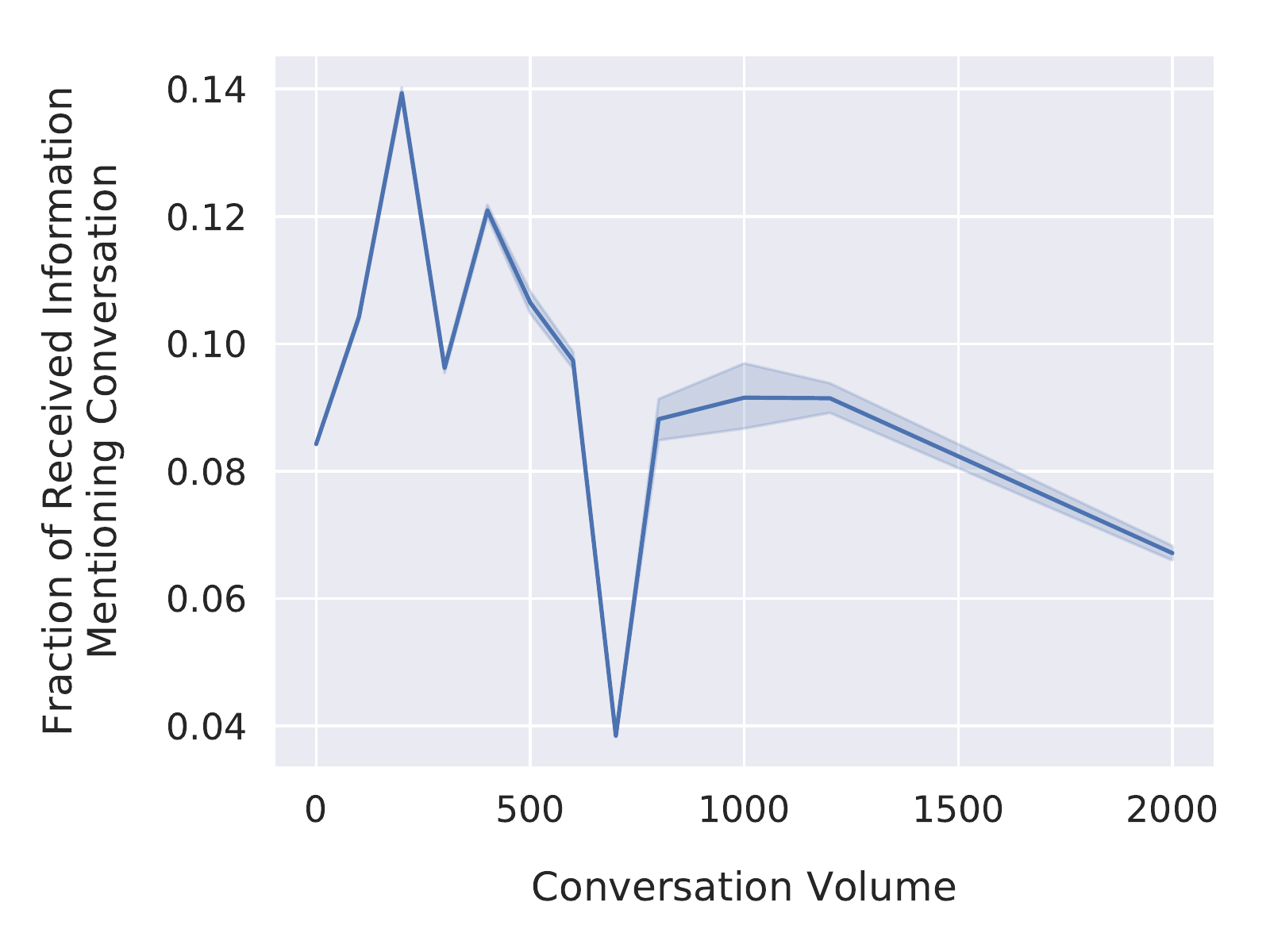}
        \caption{Fraction of received information mentioning conversation by the conversation's volume.}
        \label{fig:vol_to_convpresence_ri}
    \end{subfigure}%
    ~~
    \begin{subfigure}{0.5\textwidth}
        \centering
        \includegraphics[width=\textwidth]{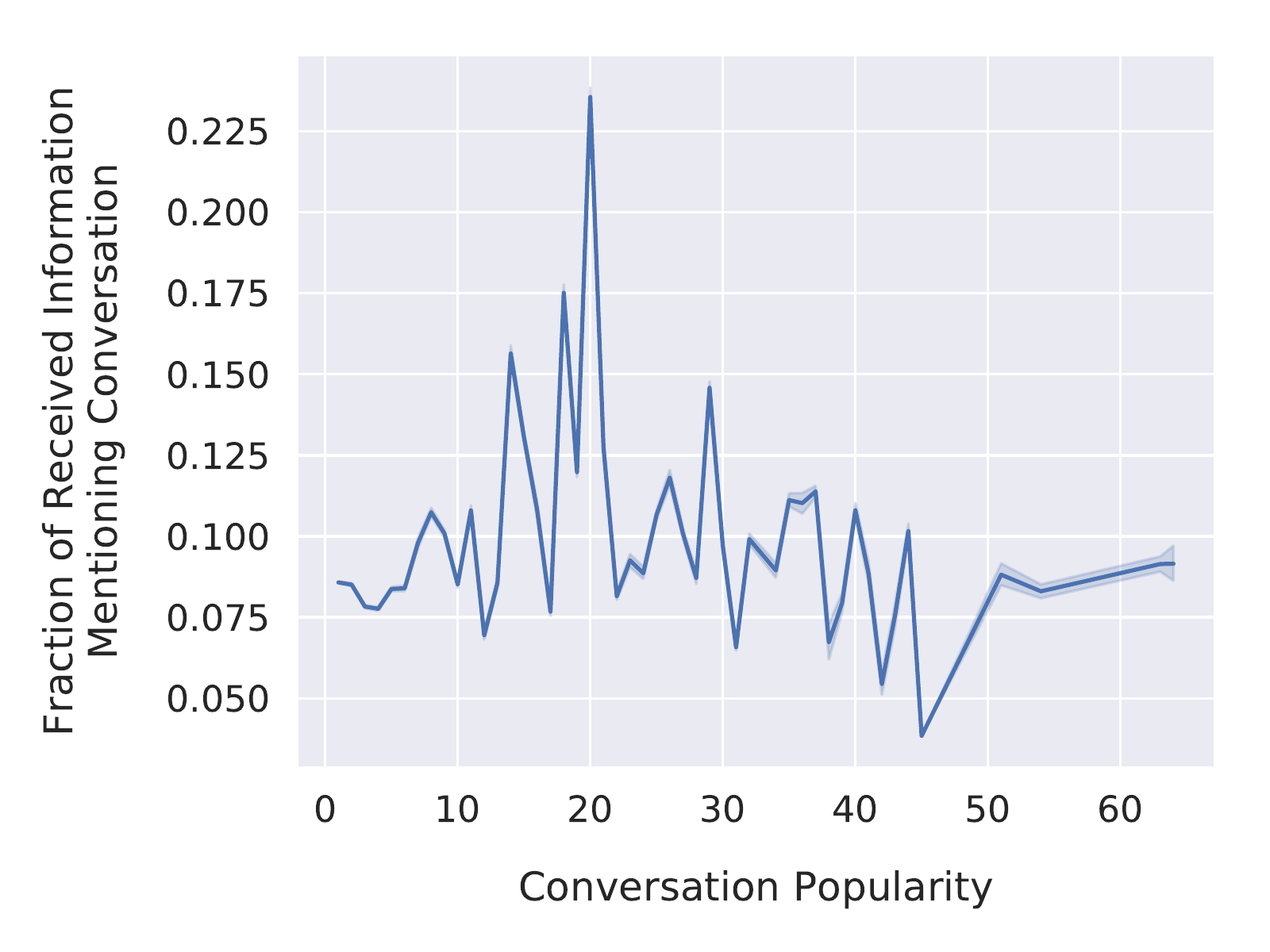}
        \caption{Fraction of influencing neighbors mentioning conversation by the conversation's global popularity.}
        \label{fig:uuc_to_convpopularity_ri}
    \end{subfigure}
    \caption{Conversation volume and popularity of messages received from individuals' local neighborhoods. Globally larger and more popular conversations were typically discussed less in local neighborhoods.}
    \label{fig:voluuc_to_convpresence_ri}
\end{figure*}

As shown in Fig. \ref{fig:volviruuc_in_out_ai_ri}, the described model of information overload replicates the analytical observation that overloaded users do receive information regarding large \citep{hodas2013friendship}. It is seen that 
overloaded users, those having more than 30 messages to consider, are unable to participate in large, popular cascades themselves in comparison to un-overloaded users, those with less than 30 messages to consider\footnote{Fig. 7 of \citep{hodas2013friendship} seems to show that highly active overloaded users are less likely to post to very large cascades than un-overloaded users. Yet, this seems to be contradicted in the text on page 230.}.
Moreover, at extreme levels of overload (50+ incoming tweets per hour), agents are so overwhelmed they are completely overloaded and unable to act. 
Additionally, we find that agents are typically able to participate in large, popular, and structurally viral cascades when they have only a few messages to process. Upon inspection of the information received by agents, we discover that cascades of globally high volume are unpopular in local neighborhoods of influence, as shown in Fig. \ref{fig:vol_to_convpresence_ri}. Conversations that have a moderate number of unique participants are more abundant in local neighborhoods of influence, as shown in  Fig. \ref{fig:uuc_to_convpopularity_ri}, but past a popularity of around 20 users, cascades quickly become less common among local neighborhoods. Therefore, this observation supports 
the similar explanation speculated in \citep{hodas2013friendship}.



\subsubsection{Contagiousness of Information}
\label{sec:hypo_info_contagion}
Finally, \citep{feng2015competing} explain that due to the information loss caused by overload, information spread dynamics do not necessarily follow those modeled in SIR models of biological contagion, contrary to prior analogies. Rather, the limiting effect of information overload leads to different epidemic thresholds. 


We observe that the probability of an agent responding to a particular conversation generally decreases past a threshold of exposures experienced by the agent, per hour. In the case of Electroneum discussions on Twitter, after a rate of around 2 exposures to a particular conversation per hour, the probability of an agent contributing to that conversation generally decreases. This evidence supports 
the findings in \citep{feng2015competing} that the mechanism of information loss under overload, may inhibit responsiveness in an individual who is excessively bombarded by notifications.

\begin{figure*}[h!]
\centering
  \includegraphics[width=0.6\textwidth]{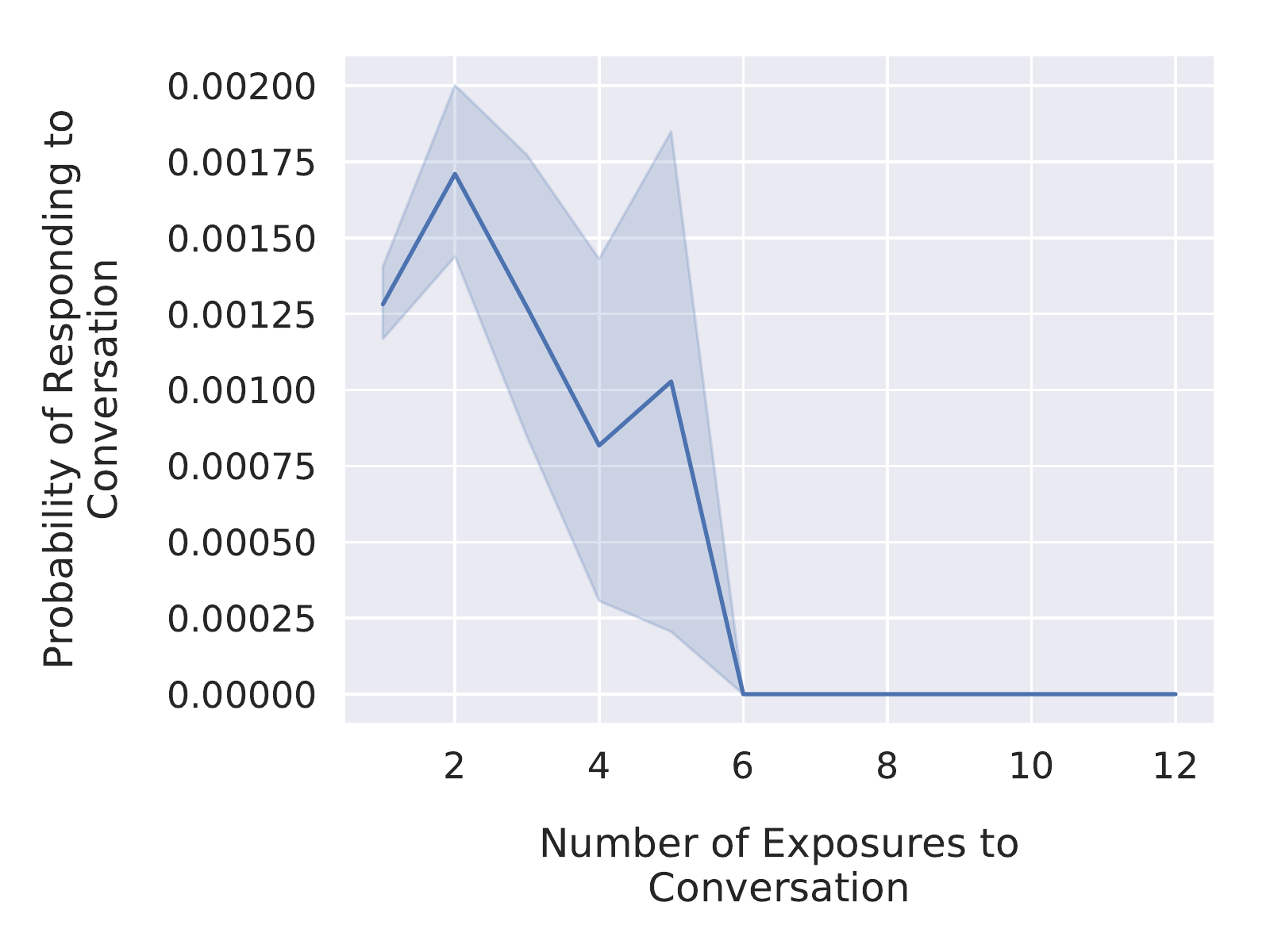}
\caption{Number of exposures to a conversation versus the probability of responding to the conversation. The probability of response declines past a threshold of exposures due to information overload.}
\label{fig:exporesprob}       
\end{figure*}




\section{Discussion and Conclusions}
From email to Instagram, asynchronous communication media have dominated our exchange of information. With exponentially higher connectivity and the increasing norm to respond later, individuals constantly find themselves with a backlog of information to respond to. Often, the rate at which information collects within the technological scaffolding of these platforms, exceeds the rate at which users can comfortably process and respond. The resulting information overload affects the dynamics of conversation and information propagation. 

In this study, we construct a mechanistic model of information overload over asynchronous communication media, informed by analytical observations on social media. This model attempts to represent the mechanism through which the subset of information received is deemed actionable under information overload. The effects of two parameters, the hourly information overload threshold, $M_{max}$, and the rate of information loss under overload, $\alpha$, on conversation characteristics, are studied through simulations of this model. The simulation-based approach provided the ability to gain useful insights into these two parameters, which would otherwise be difficult and costly to measure through direct human experimentation. Our experiments suggest that both the information overload threshold of an individual and the rate of information loss under overload play significant roles in the generation of the observable characteristics of online conversations.



Our results indicate that the amount of participation on conversations is highly sensitive at higher information overload thresholds and lower rates of information loss under overload. Conversation volume showed a similar, yet weaker relationship to these two parameters. The structure of conversations did not show any sensitivity to the information overload parameters. The information overload mechanism does not discriminate or prioritize particular users over others; prioritization is based on recency of the messages received. If instead, messages by neighbors with stronger social influence were prioritized and remained longer in the actionable information queue, then $\alpha$ and $M_{max}$, may have had an impact on virality, as they would then determine how many of the more influential neighbors (over less influential ones) would be considered for response under overload. In such a case, if more influential neighbors remained and less influential neighbors were missed due to high $\alpha$, then conversations may have had lower virality, than if $\alpha$ was low; the opposite is speculated to be true for $M_{max}$ in such a case. Romero et al. find that political discussions tend to have significantly high structural deviations from average than do other topics discussed online \citep{romero2011differences}. Assuming that the users considered by Romero et al. also experienced information overload, a prioritization of response towards certain messages based on cues other than recency, such as user familiarity, resonance of ideologies, or popularity of content, could generate sensitivity of overall conversation virality to information overload.

It is observed that the claim in \citep{li2014modeling} that responsiveness is independent of information overload threshold, only holds when it is assumed that no actionable information is lost under overload. However, our calibration results indicate that it is highly likely that there is considerable loss of actionable information due to information overload on online social media. Further, our sensitivity analyses suggest that at high rates of information loss under overload, responsiveness actually shows considerable sensitivity to the information overload threshold and is only robust for relatively high values. 

Calibration of the model to the Electroneum Twitter community shows that information overload threshold and rate of information loss under overload exist in a Pareto-front of optimal values, from mid-range information overload thresholds and low rates of information loss, to high information overload thresholds and high rates of information loss. This may indicate that several archetypes of individuals with mid to high information overload thresholds of around 20 to 30 Twitter notifications with corresponding rates of information loss per hour from 0.1 to 0.8, respectively, may exist within the studied community. Interestingly, individuals with information overload thresholds of 30 Tweets exist on this Pareto-front, corresponding to findings in \citep{gomez2014quantifying} where the responsiveness of individuals was stable up to and began to decline beyond a threshold of 30 incoming tweets an hour.

Agents of the calibrated simulations were observed to have a maximum actionable information capacity of 7 tweets an hour due to constant information overload. In other words, it was most likely that the Twitter users being simulated only considered around 7 Tweets an hour for response, while missing much of the other incoming messages. Interestingly, this observation corresponds to Miller's magical number, that individuals typically engage in cognitive tasks with $7 \pm 2$ units of information \citep{miller1956magical}. However, more recent estimates show this value to be around 4 single-item chunks of stimuli \citep{cowan2001magical}. 
We speculate whether this discrepancy is a result of the fact that the social media platform's technological scaffolding provides the extra cognitive resources necessary to have a higher processing capability than naturally possible. Alternatively, this could indicate that the information representation used, activity messages, may not represent the completely compressed form in which they are considered for response by the individuals. Studies have shown how considering uncompressed units of information may lead to the short-term memory chunk capacity to be over-estimated at 7, while actually being 3 to 4 units \citep{mathy2012s}. This could be confirmed by applying the compression strategy using Kolmogorov complexity, as suggested by Mathy \& Feldman, to test for the existence of further compressability of the activity message structure.

Similar to \citep{hodas2013friendship}, our results agree that overloaded users receive large, popular cascades of information, yet, are unable to join in these discussions. Thus we confirm the understanding that overloaded users are not good indicators of important information, as they `absorb' and mitigate the spread of large cascades. Thanks to the explanatory nature of the agent-based model, we are able to confirm the speculation that this phenomenon is caused by the local popularity of small cascades of modest global popularity and local unpopularity of larger cascades of high global popularity. 

Finally, contrary to previous research, these results confirm the findings of \citep{feng2015competing} that the spread of information is distinct from the viral spread of contagious disease. SIR models of biological contagion rely on the premise that the more an individual comes into contact with infected others, the more likely they are to become infected themselves. However, by explicitly modeling the effects of information overload, our results agree with  \citep{feng2015competing}, in that over asynchronous communication media, an individual coming into simultaneous contact with many other individuals `infected' with information may actually decrease their chance of spreading the information themselves.
Reconceptualizing information spread as distinct from biological viral spread will enable an innovative view of this process, and aid in a deeper understanding of how information is transmitted on social media.

The current model does have certain limitations. Although the rate of loss of information under overload is modeled, we do not parameterize the rate of recovery of the actionable information queue length, $M_t$. In the above experiments, we have assumed that when considering a niche online discussion forum, such as Twitter discussions of the Electroneum interest community, over the simulated time period of a month, the rate of such recovery would be negligible. Ideally, in future work, we intend to investigate the effect of modeling the recovery of information considered for response, through a parameter, say $\beta$, to counteract the effect of information loss when overloaded individuals receive information below their information overload threshold. Given the considerable match of conversation characteristics to the real-world data when the model was calibrated, we speculate that such a rate of recovery from information overload would be relatively small compared to the rate of information loss. A further limitation of this study is that we are unable to claim that external interruptions to users' activity outside of the social media platform are fully accounted for. Though an approximation of the users' typical time spent online is captured through the analysis of the activity timeseries conducted by the MACM, the model assumes that, at an hourly resolution, users receive and see the messages propagated by their neighbors.

We believe that these results provide the empirical foundations for theoretical advancement on information dynamics on asynchronous communication media. By basing our results on an explainable agent-based model, we have made it possible to track units of information as they follow the mechanisms of storage and loss of information. Analogies may be drawn between the mechanisms of the actionable information queue and concepts in the vast literature on working memory \citep{baddeley1974working,baddeley2012working,cowan2001magical,cowan2008differences,cowan2010seven,mathy2012s}. The theory of extended self \citep{clark1998extended,belk2013extended,belk2014digital,Heersmink2017,clowes2017extended} may be applied to further understand how the re-embodiment of the individual through their social media profiles could define the information dynamics between their internal working memory and their respective notification feeds, which have been modeled here. These theoretical constructs could lay the foundations for a theory of \textit{extended working memory}, an architecture defining the temporary storage of information within an individual's extended-self and its interactions with internal working memory, when engaged in asynchronous information exchange within such user-media systems.

\begin{acknowledgements}
The authors would like to acknowledge the support provided by the DARPA SocialSim program (HR001117S0018) for access to data and funding that enabled this study.
\end{acknowledgements}

%
%


\bibliographystyle{spbasic}
\bibliography{CMOT}

\end{document}